\documentclass[%
superscriptaddress,
preprint,
 amsmath,
 amssymb,
 aps,
]{revtex4-2}

\nonstopmode 

\usepackage{appendix}
\usepackage{array}
\usepackage{tikz}
\usepackage{bm} 
\usepackage{booktabs}
\usepackage{dcolumn} 
\usepackage{graphicx} 
\usepackage[hidelinks]{hyperref}
\usepackage{makecell}
\usepackage{multirow}
\usepackage{tabularray}
\usepackage{xcolor}
\usepackage{caption}
\usepackage{subcaption}

\newcommand{\highlightred}[1]{%
  \colorbox{red!20}{$\displaystyle#1$}}
\newcommand{\highlightblue}[1]{%
  \colorbox{blue!20}{$\displaystyle#1$}}

\usepackage{verbatim}

\allowdisplaybreaks

\newcommand{\ab}{\boldsymbol{a}}
\newcommand{\bb}{\boldsymbol{b}}
\newcommand{\cb}{\boldsymbol{c}}

\newcommand{\kb}{\boldsymbol{k}}

\newcommand{\pb}{\boldsymbol{p}}

\newcommand{\rb}{\boldsymbol{r}}

\newcommand{\vb}{\boldsymbol{v}}
\newcommand{\xb}{\boldsymbol{x}}

\newcommand{\Ab}{\boldsymbol{A}}
\newcommand{\Bb}{\boldsymbol{B}}

\newcommand{\Eb}{\boldsymbol{E}}

\newcommand{\Jb}{\boldsymbol{J}}

\newcommand{\Mb}{\boldsymbol{M}}

\newcommand{\Qb}{\boldsymbol{Q}}

\newcommand{\Ub}{\boldsymbol{U}}

\newcommand{\xib}{\boldsymbol{\xi}}

\newcommand{\dd}{\textrm{d}}
\newcommand{\de}{\partial}
\newcommand{\Nabla}{\boldsymbol{\nabla}}
\newcommand{\eps}{\varepsilon}

\newcommand{\wh}{\widehat}

\newcommand{\vbgal}{\vb_{\textrm{gal}}}

\newcommand{\dt}{\Delta t}

\newcommand{\spectral}[1]{\widehat{#1}}

\begin{document}

\preprint{APS/123-QED}

\title{
PIC-JR$_{ho}$m: a pseudo-spectral Particle-In-Cell formulation \\
with arbitrary charge and current densities time dependencies \\
for the modeling of relativistic plasmas}
 
\author{Olga Shapoval}
\affiliation{Lawrence Berkeley National Laboratory, Berkeley, CA 94720, USA}

\author{Edoardo Zoni}
\altaffiliation[Current affiliation: ]{Centro Euro-Mediterraneo sui Cambiamenti Climatici, 40127 Bologna, Italy}
\affiliation{Lawrence Berkeley National Laboratory, Berkeley, CA 94720, USA}


\author{Remi Lehe}
\affiliation{Lawrence Berkeley National Laboratory, Berkeley, CA 94720, USA}

\author{Maxence Thévenet}
\affiliation{Deutsches Elektronen-Synchrotron DESY, Notkestr. 85, 22607 Hamburg, Germany}

\author{Jean-Luc Vay}
\email[Corresponding author: ]{jlvay@lbl.gov}
\affiliation{Lawrence Berkeley National Laboratory, Berkeley, CA 94720, USA}

\date{\today}

\begin{abstract}
This paper introduces a novel formulation of the Particle-In-Cell (PIC) method for the modeling of relativistic plasmas, which leverages the ability of the Pseudo-Spectral Analytical Time-Domain solver (PSATD) to handle arbitrary time dependencies of the charge and current densities during one PIC cycle. 
The new formulation is applied to a modified set of Maxwell's equations that was proposed earlier in the context of divergence cleaning, and to recently proposed extensions of the PSATD-PIC algorithm.
Detailed analysis and testings revealed that, under some condition, the new formulation can expand the range of numerical parameters under which PIC simulations are stable and accurate when modeling relativistic plasmas such as, e.g., plasma-based particle accelerators. 

\end{abstract}

\maketitle

\section{
\label{sec:introduction}
Introduction}
Simulations of relativistic plasmas often rely on the electromagnetic particle-in-cell (PIC) method~\cite{BunemanJCP1980,DawsonRMP83,Birdsalllangdon}, with variations of the method that have been proposed and are chosen based on the application. For the modeling of plasma-based accelerators~\cite{Tajimaprl79,ChenPRL1985}, a variation that has gained in popularity uses the ``infinite-order'' (in space and time) Pseudo-Spectral Analytical Time-Domain (PSATD) method~\cite{Haber1973,VayJCP13}, instead of the (almost universally adopted) second-order (in space and time) Finite-Difference Time-Domain (FDTD) ``Yee'' method~\cite{Yee}, to solve Maxwell's equations at discrete points in space and time. In contrast to the Yee solver, the PSATD solver offers no numerical dispersion and no Courant condition on the field solve. Extensions of the PSATD PIC method includes the use of finite-order spatial stencils~\cite{VincentiCPC2016,VincentiCPC2018}, alternating nodal-staggered representations of the field quantities during one PIC loop~\cite{ZoniCPC2022}, time-averaging of the fields gathered onto the particles~\cite{ShapovalPRE2021}, and integration of the equations in a Galilean frame moving at a given velocity~\cite{LehePRE2016,KirchenPoP2016}. The combination of the Galilean PSATD PIC (also labeled as Galilean PIC for convenience) method with the other extensions has led to stable modeling of plasma accelerators,  
free of the numerical Cherenkov instability (NCI)~\cite{Godfreyjcp74} 
when using the Lorentz boosted frame method to speed up simulations~\cite{Vayprl07}. In some cases, however, the method, which relies on the user setting a predefined Galilean velocity, can become inaccurate when it cannot be assumed that the local plasma velocity is close to that predefined velocity.
As a possible remedy, this paper introduces and starts exploring a novel formulation of the PIC algorithm where the standard assumption that the current density that is produced by the particles is constant over a time step is relaxed. 

The reminder of the paper is organized as follows. The formulation of the novel 
algorithm is derived first, followed by the presentation of its finite-order stencil, alternating nodal-staggered and time-averaged extensions. The connection between the new algorithm and the Galilean PIC formulation is discussed next. The effectiveness of the new algorithm at mitigating the NCI is then explored theoretically and numerically on a simple uniform plasma case. Finally, the new scheme is tested in simulations of laser-plasma accelerators in a Lorentz boosted frame.

\section{
\label{sec:psatd_equations}
Novel 
PIC-JR\texorpdfstring{$_{ho}$\MakeLowercase{m}}{hom}
algorithm}

\subsection{
\label{subsec:psatd_equations_derivation}
Presentation of the algorithm}

 The following modified system of Maxwell's equations is considered 
\begin{subequations}
\label{Maxwell}
\begin{align}
\label{dE_dt}
\frac{\de\Eb}{\de{t}} & = c^2\Nabla\times\Bb-\frac{\Jb}{\eps_0}+c^2\Nabla{F} \,, \\[5pt]
\label{dB_dt}
\frac{\de\Bb}{\de{t}} & = -\Nabla\times\Eb \,, \\[5pt] 
\label{dF_dt}
\frac{\de F}{\de{t}} & = \Nabla\cdot\Eb-\frac{\rho}{\eps_0} \,. 
\end{align}
\end{subequations}
In addition to the usual Maxwell-Faraday and Amp\`ere-Maxwell equations, the system contains an extra equation for 
the scalar field $F$, which propagates deviations to Gauss' law. (Note that, in the case where Gauss' law is verified in the PIC simulation, Eq.~\eqref{dF_dt} leads to $F=0$, and Eqs.~\eqref{dE_dt},\eqref{dB_dt} reduce to the standard Maxwell's equations.) These additional terms were introduced in~\cite{Vay1996} from the potential formulation in the Lorentz gauge 
and used as a propagative divergence cleaning procedure, as an alternate to the Langdon-Marder~\cite{Langdoncpc92} or Marder~\cite{Marderjcp87} diffusive ones. This type of divergence cleaning was also proposed independently and analyzed more formally in~\cite{Munz2000}. A connection to the formulation of Eqs.~\eqref{Maxwell} in potential form, derived more formally than in~\cite{Vay1996}, is instructive and given in Appendix~\ref{sec:Appendix_derivation_from_potentials}.

While the abovementioned earlier work~\cite{Vay1996,Munz2000} considered this formulation in the context of the standard PIC method using FDTD discretization of Eqs.~\eqref{Maxwell},  
this article focuses on the PSATD~\citep{Haber1973,BunemanJCP1980,Vay2013} discretization of Eqs.~\eqref{Maxwell}, where the equations are integrated analytically over one timestep, in Fourier space. The expression of~\eqref{Maxwell} in Fourier space reads
\begin{subequations}
\label{Maxwell_k}
\begin{align}
\label{dE_dt_k}
\frac{\de\wh\Eb}{\de{t}} & = {i}c^2\kb\times\wh\Bb-\frac{\wh\Jb}{\eps_0}+{i}c^2\wh{F}\kb \,, \\[5pt]
\label{dB_dt_k}
\frac{\de\wh\Bb}{\de{t}} & = -i\kb\times\wh\Eb \,, \\[5pt] 
\label{dF_dt_k}
\frac{\de\wh F}{\de{t}} & = i\kb\cdot\wh\Eb-\frac{\wh\rho}{\eps_0} \,, 
\end{align}
\end{subequations}
where $\wh f$ denotes the Fourier transform of function $f$.
The analytical integration of Eqs~\eqref{Maxwell_k} in time requires an assumption on the time dependency of the current and charge densities $\wh\Jb$ and $\wh\rho$ over the integration interval, i.e., over a timestep that goes from $t=n\Delta t$ to $t=(n+1)\Delta t$. In the standard PSATD algorithm \cite{Haber1973,VayJCP13}, $\wh\Jb$ is assumed to constant in time, and $\wh\rho$ is assumed to be linear in time, within a given timestep $\Delta t$.

\begin{figure}
\includegraphics[width=0.96\linewidth]{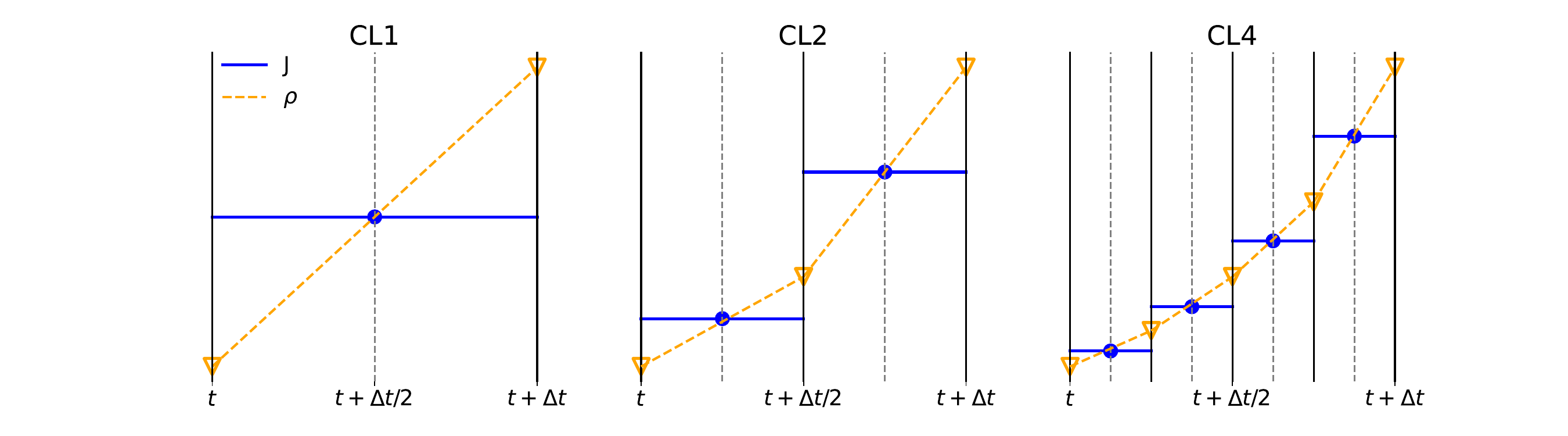}
\includegraphics[width=0.96\linewidth]{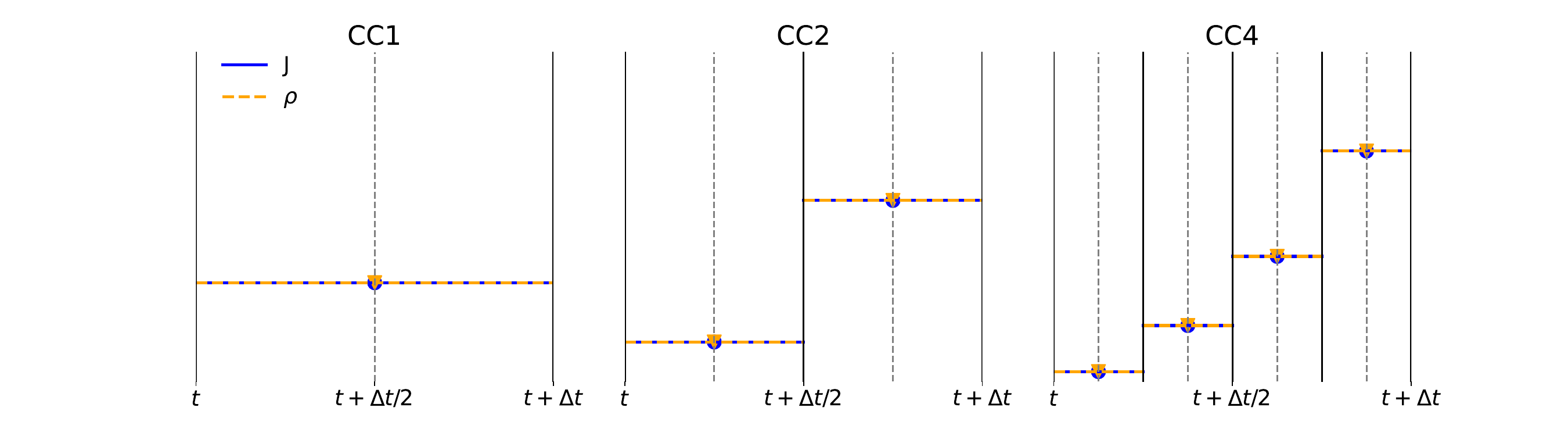}
\includegraphics[width=0.96\linewidth]{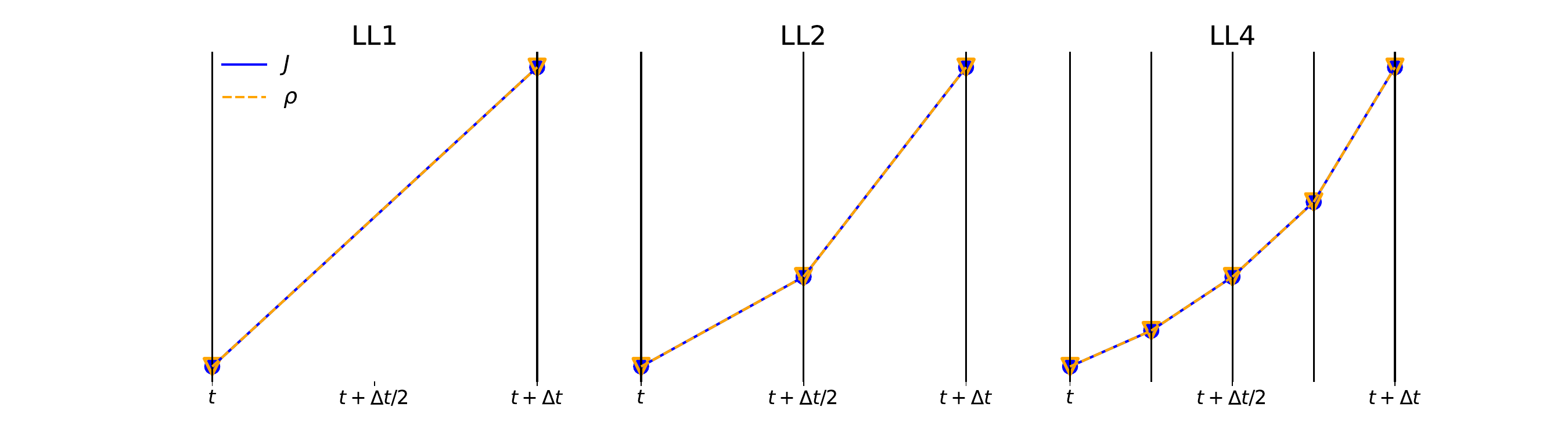}
\includegraphics[width=0.96\linewidth]{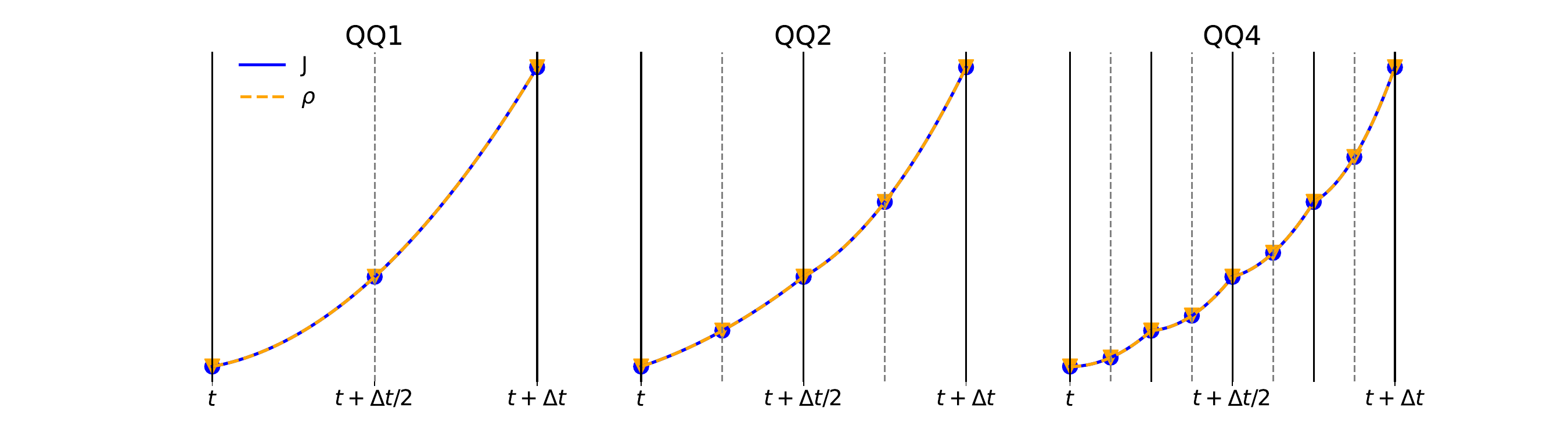}
\caption{Diagrams illustrating various time dependencies of the current density $\Jb$ and charge density $\rho$ for constant/linear (CL), both constant (CC), linear (LL) and quadratic (QQ) dependencies with $m$ subintervals: (first column) $m=1$, (second) $m=2$ and (third) $m=4$. CL1 corresponds to the standard PSATD PIC method. The triangle and circle glyphs represent the times at which the macroparticles deposit $\rho$ and $\Jb$ on the grid, respectively. The dashed and solid lines represent the assumed time dependency of $\rho$ and $\Jb$ within one timestep, when integrating the Maxwell equations analytically.}
\label{fig:P2PIC}
\end{figure}

This paper considers more general time dependencies for $\wh\Jb$ and $\wh\rho$ within one timestep, which is divided into $m$ subintervals of equal size $\delta t = \Delta t/m$. During these subintervals, $\wh\Jb$ and $\wh\rho$ are considered to be either piecewise constant, piecewise linear, or piecewise quadratic in time. This is illustrated in Fig.~\ref{fig:P2PIC}. In the rest of this paper, the notation ``PIC-JR$_{ho}$m'' is used, where J and R$_{ho}$ (J,R$_{ho}$ $\in \{$C (constant), L (linear), Q (quadratic)$\}$) indicate the (piecewise) time dependency of the current density $\wh \Jb$ and charge density $\wh \rho$, respectively, and $m$ is the number of subintervals. For example, ``PIC-LL2`` refers to the novel PIC algorithm with linear time dependency of both $\wh \Jb$ and $\rho$ and 2 subintervals. Note that, in this notation, ``PIC-CL1`` refers to the standard PSATD PIC algorithm \cite{Vay2013}, where $\wh\Jb$ is constant and $\wh \rho$ is linear in time over one time step.

More specifically:
\begin{itemize}
\item When $\wh\rho(t)$ is assumed to be \textbf{piecewise constant}: macroparticles deposit their charge density in the middle of each time subinterval, i.e., at $t_{n+(\ell+1/2)/m} \equiv n\Delta t + (\ell+1/2)\delta t$ with $\ell\in[0,m-1]$, and $\wh\rho$ is then assumed to be constant in each subinterval: 
\[ \wh\rho(t) = \rho^{n+(\ell+1/2)/m},\quad t\in[ n\Delta t + \ell\delta t, n\Delta t + (\ell+1)\delta t]. \]

\item When $\wh\rho(t)$ is assumed to be \textbf{piecewise linear}: macroparticles deposit their charge density at the edge of each time subinterval, i.e., at $t_{n+\ell/m} \equiv n\Delta t + \ell\delta t$, with $\ell\in[0,m]$, and $\wh\rho$ is then assumed to be linear in each subinterval: 
\begin{align*}
\wh\rho(t) =& \frac{\wh\rho^{n+(\ell+1)/m}-\wh\rho^{n+\ell/m}}{\delta t}(t-t_{n+(\ell+1/2)/m}) + \frac{\wh\rho^{n+(\ell+1)/m}+\wh\rho^{n+\ell/m}}{2},\\
&t\in[ n\Delta t + \ell\delta t, n\Delta t + (\ell+1)\delta t].
\end{align*}

\item When $\wh\rho(t)$ is assumed to be \textbf{piecewise quadratic}: macroparticles deposit their charge density at the middle \emph{and} edge of each time subinterval, i.e., at $t_{n+(\ell+1/2)/m}$ with $\ell\in[0,m-1]$ and $t_{n+\ell/m}$ with $\ell\in[0,m]$. $\wh\rho$ is then assumed to be quadratic in each subinterval: 
\begin{align*}
\wh\rho(t) =& \frac{2(\rho^{n+(\ell+1)/m}-2\wh\rho^{n+(\ell+1/2)/m}+\rho^{n+\ell/m})}{\delta t^2}(t-t_{n+(\ell+1/2)/m})^2 \\
&+ \frac{\wh\rho^{n+(\ell+1)/m}-\wh\rho^{n+\ell/m}}{\delta t}(t-t_{n+(\ell+1/2)/m}) + \rho^{n+(\ell+1/2)/m}, \\
&t\in[ n\Delta t + \ell\delta t, n\Delta t + (\ell+1)\delta t],
\end{align*}
\end{itemize}
with similar definitions for $\wh\Jb$, when $\wh\Jb(t)$ is assumed to be piecewise constant, piecewise linear, or piecewise quadratic respectively.

Overall, the time dependency of $\wh\Jb$ and $\wh\rho$ can thus be expressed, for $t\in[ n\Delta t + \ell\delta t, n\Delta t + (\ell+1)\delta t]$, with $\ell\in [0, m-1]$, as:
\begin{subequations}
\label{Jrhotime}
\begin{align}
\label{eq:j_func}
\wh\Jb(t) & = \frac{2\ab_{\Jb}^{\tau}}{\delta t^2}(t-t_{n+(\ell+1/2)/m})^2+\frac{\bb_{\Jb}^{\tau}}{\delta t}(t-t_{n+(\ell+1/2)/m})+\cb_{\Jb}^{\tau} \,, \\[5pt]
\label{eq:rho_func}
\wh\rho(t) & = \frac{2a_{\rho}^{\tau}}{\delta t^2}(t-t_{n+(\ell+1/2)m})^2+\frac{b_{\rho}^{\tau}}{\delta t}(t-t_{n+(\ell+1/2)m})+c_{\rho}^{\tau} \,,
\end{align}
\end{subequations}
where the coefficients of the polynomials are given in Table~\ref{table:JR_polynom_coefs}.

It is important to note that the particles' momenta are not updated during one time step, i.e., the proposed scheme does \emph{not} involve subcycling of the macroparticles motion. As in standard PSATD PIC, 
macroparticles move in straight line from their known position at $t_n=n\Delta t$ to time $t$, using their known momentum at $t_{n+1/2}$:
\[ \boldsymbol{x}(t) = \boldsymbol{x}^n + \frac{\boldsymbol{p}^{n+1/2}}{m\sqrt{1+ (\boldsymbol{p}^{n+1/2}/mc)^2}}(t-t_n) \]
where $\boldsymbol{x}^n$ and $\boldsymbol{p}^{n+1/2}$ follow the standard leap-frog time stepping that is commonly used in PIC simulations.
Thus, here, even though the charge and current density may be deposited several times per timestep $\Delta t$, the macroparticles' momentum $\boldsymbol{p}$ is only updated once per timestep, and therefore the fields $\boldsymbol{E}$ and $\boldsymbol{B}$ are gathered onto macroparticles to update $\boldsymbol{p}$ only once per timestep also. 

\begin{table}[t]
\setlength{\tabcolsep}{8pt}
\begin{tabular}{cccc}
\toprule
\multirow{2}{*}{\textbf{PC}}& \multicolumn{3}{c}{Time dependency of $\wh \Jb$ or $\wh \rho$} \\
\cline{2-4}
& constant  ($\tau=0$) & linear  ($\tau=1$) & quadratic  ($\tau=2$) \\ 
\midrule
${{\ab}_{J}^{\tau}}$ & 0 & 0 & $\wh \Jb^{n+(\ell+1)/m} -2\wh \Jb^{n+(\ell+1/2)/m}+\wh \Jb^{n+\ell/m}$ \\ 
${{\bb}_{J}^{\tau}}$ & 0 & $\wh \Jb^{n+(\ell+1)/m}-\wh \Jb^{n+\ell/m}$ & $\wh \Jb^{n+(\ell+1)/m}-\wh \Jb^{n+\ell/m}$ \\
${{\cb}_{J}^{ \tau}}$ & $\wh \Jb^{n+(\ell+1/2)/m}$ & $(\wh \Jb^{n+(\ell+1)/m}+\wh \Jb^{n+\ell/m})/2$ & $\wh \Jb^{n+(\ell+1/2)/m}$ \\
${{a}_{\wh \rho}^{\tau}}$ & 0 & 0 & $\wh \rho^{n+(\ell+1)/m} -2\wh \rho^{n+(\ell+1/2)m}+\wh \rho^{n+\ell/m}$ \\ 
${{b}_{\wh \rho}^{\tau}}$ & 0 & $\wh \rho^{n+(\ell+1)/m}-\wh \rho^{n+\ell/m}$ & $\wh \rho^{n+(\ell+1)/m}-\wh \rho^{n+\ell/m}$ \\
${{c}_{\wh \rho}^{ \tau}}$ & $\wh \rho^{n+(\ell+1/2)/m}$ & $(\wh \rho^{n+(\ell+1)/m}+\wh \rho^{n+\ell/m})/2$ & $\wh \rho^{n+(\ell+1/2)m}$ \\
\bottomrule
\end{tabular}
\caption{\label{table:JR_polynom_coefs}Polynomial coefficients (\textbf{PC}), based on the time dependency of the current and charge densities $\wh \Jb$ and $\wh \rho$ over $\ell$-th time subinterval $[n \Delta t + \ell \delta t,n \Delta t + (\ell+1) \delta t]$.}
\end{table}


%

Using the piecewise definition of $\wh\rho$ and $\wh\Jb$ given in Eqs.~\eqref{Jrhotime}, Eqs.~\eqref{Maxwell_k} can be integrated analytically over one timestep $\Delta t$, i.e., from $t=n\Delta t$ to $t=(n+1)\Delta t$. In practice, this is done by sequentially integrating these equations over each subinterval $\ell \in [0,m-1]$:
\begin{subequations}
\label{Maxwell_discrete}
\begin{align}
\wh\Eb^{n+(\ell+1)/m} & = C\wh\Eb^{n+\ell/m}+ic^2\frac{S}{ck}\kb\times\wh\Bb^{n+\ell/m}+ic^2\frac{S}{ck}\wh{F}^{n+\ell/m}\kb+\frac{1}{\eps_0 ck}\left(Y_3\ab_J+Y_2\bb_J-S\cb_J\right) \nonumber \\
& \quad +\frac{ic^2}{\eps_0 c^2k^2}\left({Y_1}a_{\rho}-Y_{5}b_{\rho}-Y_{4}c_{\rho}\right)\kb \,,  \\[5pt] 
\wh \Bb^{n+(\ell+1)/m} & = C\wh \Bb^{n+\ell/m}-i\frac{S}{ck }\kb\times\wh\Eb^{n+\ell/m}-\frac{i}{\eps_0 c^2k^2}\kb\times\left(Y_1\ab_J-Y_5\bb_J-Y_4\cb_J\right) \,, \\[5pt]
\wh F^{n+(\ell+1)/m} & = C \wh F^{n+\ell/m}+i\frac{S}{ck}\kb\cdot\wh \Eb^{n+\ell/m}+\frac{i}{\eps_0 c^2k^2}\kb\cdot\left(Y_1\ab_J-Y_5\bb_J-Y_4\cb_J\right) \nonumber \\
& \quad +\frac{1}{\eps_0 ck}\left({Y_3}a_{\rho}+{Y_2}b_{\rho}-Sc_{\rho}\right) \, 
\end{align}
\end{subequations}
where
\begin{align}
\begin{split}
\label{coefs_x1_x2}
C &= \cos(ck\delta t), \ S = \sin(ck\delta t), \\
Y_1 & = \frac{(1-C)(8-c^2k^2\delta t^2)-4Sck\delta t}{2 c^2 k^2 \delta t^2},\\
Y_2 & = \frac{2(C-1)+ S ck\delta t }{2 ck\delta t}, \\
Y_3 & = \frac{S(8- c^2k^2\delta t^2 ) - 4ck\delta t(1+C)}{2c^2 k^2 \delta t^2},\\
Y_4 &= (1-C), \ Y_5 = \frac{(1+C) ck\delta t - 2S}{2ck \delta t}. \\
\end{split}
\end{align}

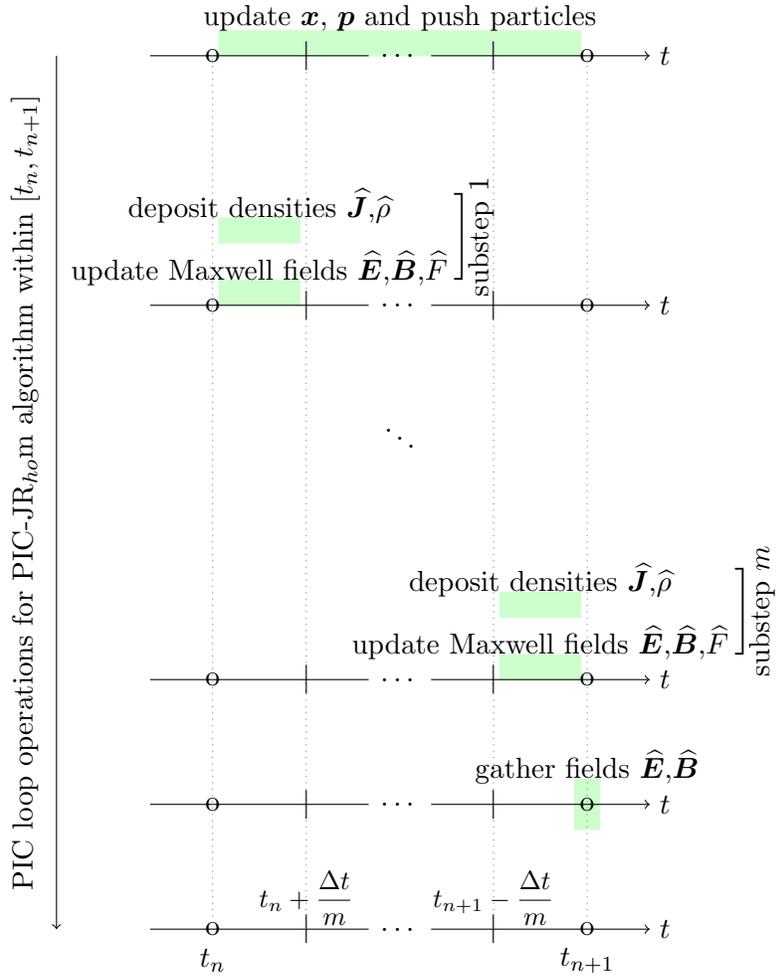
\begin{figure}[htb!]
\centering
\begin{tikzpicture}[scale=0.83]
    \def \y {0}
    \filldraw[fill=green!20, draw=green!20] (0.1,\y) rectangle (6-0.1,\y+0.4);
    \node at (3,\y+0.6) {update $\xb$, $\pb$ and push particles};
    \def \xmin {0}
    \def \xmax {1.5}
    \def \xmid {\xmin + \xmax/2 - \xmin/2}
    \def \y {-3}
    \filldraw[fill=green!20, draw=green!20] (\xmin+0.1,\y) rectangle (\xmax-0.1,\y+0.4);
    \node at (\xmid,\y+0.6) {deposit densities $\wh \Jb$,$\wh \rho$};
    \def \y {-4}
    \filldraw[fill=green!20, draw=green!20] (\xmin+0.1,\y) rectangle (\xmax-0.1,\y+0.4);
    \node at (\xmid,\y+0.6) {update Maxwell fields $\wh \Eb$,$\wh \Bb$,$\wh F$};
    \node at (4,-2.9) {$\Bigg]$};
    \node[rotate=90] at (4+0.3,-2.9) {substep $1$};
    \node at (3,-6) {$\ddots$};
    \def \xmin {4.5}
    \def \xmax {6}
    \def \xmid {\xmin + \xmax/2 - \xmin/2}
    \def \y {-9}
    \filldraw[fill=green!20, draw=green!20] (\xmin+0.1,\y) rectangle (\xmax-0.1,\y+0.4);
    \node at (\xmid,\y+0.6) {deposit densities $\wh \Jb$,$\wh \rho$};
    \def \y {-10}
    \filldraw[fill=green!20, draw=green!20] (\xmin+0.1,\y) rectangle (\xmax-0.1,\y+0.4);
    \node at (\xmid,\y+0.6) {update Maxwell fields $\wh \Eb$,$\wh \Bb$,$\wh F$};
    \node at (8.5,-8.9) {$\Bigg]$};
    \node[rotate=90] at (8.5+0.3,-8.9) {substep $m$};
    \def \xmid {6}
    \def \y {-12}
    \filldraw[fill=green!20, draw=green!20] (\xmid-0.2,\y-0.4) rectangle (\xmid+0.2,\y+0.4);
    \node at (\xmid,\y+0.6) {gather fields $\wh \Eb$,$\wh \Bb$};
    \def \xmin {0}
    \def \xmax {6}
    \foreach \y in {0,-4,-10,-12,-14} \node at (\xmin,\y) {o};
    \foreach \y in {0,-4,-10,-12,-14} \node at (\xmin+1.5,\y) {$\vert$};
    \foreach \y in {0,-4,-10,-12,-14} \node at (\xmax-1.5,\y) {$\vert$};
    \foreach \y in {0,-4,-10,-12,-14} \node at (\xmax,\y) {o};
    \def \y {-14}
    \node at (\xmin,\y-0.5) {$t_n$};
    \node at (\xmin+1.5,\y+0.5) {\footnotesize $t_n+\dfrac{\Delta t}{m}$};
    \node at (\xmax-1.5,\y+0.5) {\footnotesize $t_{n+1}-\dfrac{\Delta t}{m}$};
    \node at (\xmax,\y-0.5) {$t_{n+1}$};
    \foreach \y in {0,-4,-10,-12,-14} \draw (-1,\y) -- (0,\y); 
    \foreach \y in {0,-4,-10,-12,-14} \draw (0,\y) -- (2.5,\y);
    \foreach \y in {0,-4,-10,-12,-14} \node at (3,\y) {$\dots$};
    \foreach \y in {0,-4,-10,-12,-14} \draw (3.5,\y) -- (6,\y);
    \foreach \y in {0,-4,-10,-12,-14} \draw[->] (6,\y) -- (7,\y); 
    \foreach \y in {0,-4,-10,-12,-14} \node[right] at (7,\y) {$t$};
    \def \ymin {-14}
    \def \ymax {0}
    \def \xmin {0}
    \def \xmax {6}
    \draw[dotted, color=gray] (\xmin    ,\ymin) -- (\xmin    ,\ymax);
    \draw[dotted, color=gray] (\xmin+1.5,\ymin) -- (\xmin+1.5,\ymax);
    \draw[dotted, color=gray] (\xmax-1.5,\ymin) -- (\xmax-1.5,\ymax);
    \draw[dotted, color=gray] (\xmax    ,\ymin) -- (\xmax    ,\ymax);
    \def \y {-14}
    \draw[->] (-2.5,0) -- (-2.5,\y);
    \node[rotate=90] at (-2.5-0.5,\y/2) {PIC loop operations for PIC-JR$_{ho}$m algorithm within $[t_n,t_{n+1}]$};
\end{tikzpicture}
\caption{Diagram of the PIC-JR$_{ho}$m algorithm.}
\label{fig:pic_loop}
\end{figure}

The steps of the novel PIC-JR$_{ho}$m cycle with sub-timestepping are summarized in the diagram shown in Figure~\ref{fig:pic_loop}.

\subsection{
\label{subsec:psatd_JRm_extensions}
Extensions}

As shown in Refs.~\cite{VincentiCPC2016, VincentiCPC2018, ZoniCPC2022,ShapovalPRE2021}, the PSATD PIC algorithm can be extended to (a) arbitrary-order spatial stencils, (b) a scheme that alternates between nodal and staggered representations of the field components on the simulation grid, and (c) a scheme that averages the fields to be gathered over one timestep. Such extensions are presented in the next sections for the PSATD PIC-JR$_{ho}$m algorithm.

\subsubsection{
\label{subsec:psatd_finiteorder}
Extension to finite-order stencils}

When using domain decomposition to run PSATD PIC methods on parallel computers, it is advantageous to alter the wave vector in the Fourier representation of the equations so as to emulate a finite-difference approximation of the spatial derivatives at a finite order $p$, since this enhances the locality of the field solvers and thus reduces the required number of guard cells around each subdomain~\cite{VincentiCPC2016,VincentiCPC2018}. 
The modified $[k^p_u]$ at order $p$ along the direction $u={x,y,z}$ are then given by
\begin{subequations}
\begin{align}
    [k^p_u]_{\textrm{nodal}} &= \sum_{j=1}^{p/2}{[\alpha_j^p]_{\textrm{nodal}} \frac{\sin(k_u j \Delta u)}{j \Delta u}}, \ \  u={x,y,z},\\ 
    [k^p_u]_{\textrm{staggered}} &= \sum_{j=1}^{p/2}{[\alpha_j^p]_{\textrm{staggered}} \frac{\sin(k_u (j-1/2) \Delta u)}{(j-1/2) \Delta u}}, \ \  u={x,y,z}, 
\end{align}    
\end{subequations}
for a nodal and staggered representation, respectively, with the following Fornberg coefficients~\cite{Fornberg1990}:
\begin{subequations}
\begin{align}
    [\alpha_j^p]_{\textrm{nodal}} &= (-1)^{j+1} \frac{2 [(p/2)!]^2}{(p/2-j)!(p/2+j)!}, \\
    [\alpha_j^p]_{\textrm{staggered}} &= (-1)^{j+1} \left[\frac{p!}{2^p (p/2)!}\right]^2 \frac{4}{(2j-1)(p/2-j)!(p/2+j-1)!}.
\end{align}    
\end{subequations}



These modified wave numbers can be readily used with the PIC-JR$_{ho}$m algorithm to limit the number of guard cells and enable efficient parallel simulations, just as with other flavors of PSATD PIC algorithms~\cite{VincentiCPC2016,VincentiCPC2018}.

\subsubsection{
\label{subsec:psatd_alternating_nodalstaggered}
Extension to alternating nodal-staggered grids}

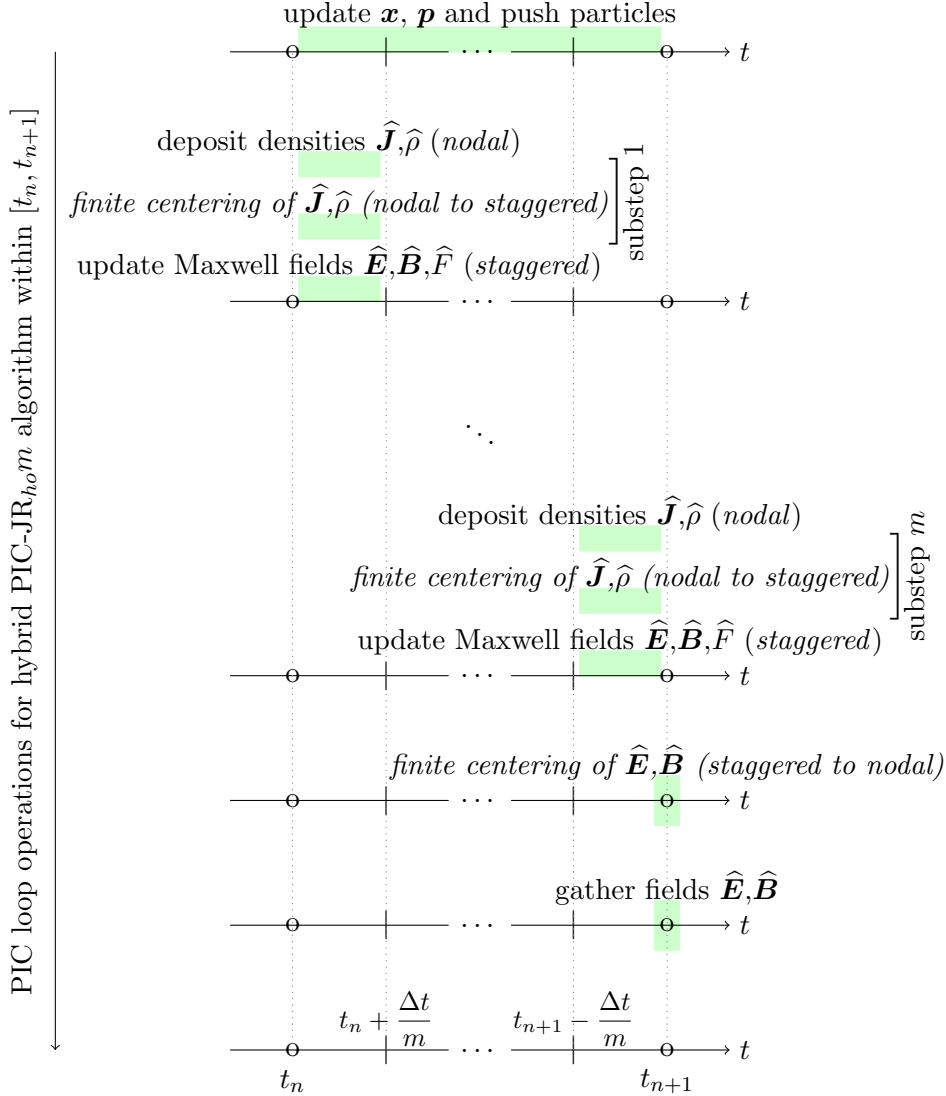
\begin{figure}[htb!]
\centering
\begin{tikzpicture}[scale=0.83]
    \def \y {0}
    \filldraw[fill=green!20, draw=green!20] (0.1,\y) rectangle (6-0.1,\y+0.4);
    \node at (3,\y+0.6) {update $\xb$, $\pb$ and push particles};
    \def \xmin {0}
    \def \xmax {1.5}
    \def \xmid {\xmin + \xmax/2 - \xmin/2}
    \def \y {-2}
    \filldraw[fill=green!20, draw=green!20] (\xmin+0.1,\y) rectangle (\xmax-0.1,\y+0.4);
    \node at (\xmid,\y+0.6) {deposit densities $\wh \Jb$,$\wh \rho$ (\textit{nodal})};
    \def \y {-3}
    \filldraw[fill=green!20, draw=green!20] (\xmin+0.1,\y) rectangle (\xmax-0.1,\y+0.4);
    \node at (\xmid,\y+0.6) {\it finite centering of $\wh \Jb$,$\wh \rho$ (nodal to staggered)};
    \def \y {-4}
    \filldraw[fill=green!20, draw=green!20] (\xmin+0.1,\y) rectangle (\xmax-0.1,\y+0.4);
    \node at (\xmid,\y+0.6) {update Maxwell fields $\wh \Eb$,$\wh \Bb$,$\wh F$ (\textit{staggered})};
    \node at (5.2,-2.4) {$\Bigg]$};
    \node[rotate=90] at (5.2+0.3,-2.4) {substep $1$};
    \node at (3,-6) {$\ddots$};
    \def \xmin {4.5}
    \def \xmax {6}
    \def \xmid {\xmin + \xmax/2 - \xmin/2}
    \def \y {-8}
    \filldraw[fill=green!20, draw=green!20] (\xmin+0.1,\y) rectangle (\xmax-0.1,\y+0.4);
    \node at (\xmid,\y+0.6) {deposit densities $\wh \Jb$,$\wh \rho$ (\textit{nodal})};
    \def \y {-9}
    \filldraw[fill=green!20, draw=green!20] (\xmin+0.1,\y) rectangle (\xmax-0.1,\y+0.4);
    \node at (\xmid,\y+0.6) {\it finite centering of $\wh \Jb$,$\wh \rho$ (nodal to staggered)};
    \def \y {-10}
    \filldraw[fill=green!20, draw=green!20] (\xmin+0.1,\y) rectangle (\xmax-0.1,\y+0.4);
    \node at (\xmid,\y+0.6) {update Maxwell fields $\wh \Eb$,$\wh \Bb$,$\wh F$ (\textit{staggered})};
    \node at (9.7,-8.4) {$\Bigg]$};
    \node[rotate=90] at (9.7+0.3,-8.4) {substep $m$};
    \def \xmid {6}
    \def \y {-12}
    \filldraw[fill=green!20, draw=green!20] (\xmid-0.2,\y-0.4) rectangle (\xmid+0.2,\y+0.4);
    \node at (\xmid,\y+0.6) {\it finite centering of $\wh \Eb$,$\wh \Bb$ (staggered to nodal)};
    \def \xmid {6}
    \def \y {-14}
    \filldraw[fill=green!20, draw=green!20] (\xmid-0.2,\y-0.4) rectangle (\xmid+0.2,\y+0.4);
    \node at (\xmid,\y+0.6) {gather fields $\wh \Eb$,$\wh \Bb$};
    \def \xmin {0}
    \def \xmax {6}
    \foreach \y in {0,-4,-10,-12,-14,-16} \node at (\xmin,\y) {o};
    \foreach \y in {0,-4,-10,-12,-14,-16} \node at (\xmin+1.5,\y) {$\vert$};
    \foreach \y in {0,-4,-10,-12,-14,-16} \node at (\xmax-1.5,\y) {$\vert$};
    \foreach \y in {0,-4,-10,-12,-14,-16} \node at (\xmax,\y) {o};
    \def \y {-16}
    \node at (\xmin,\y-0.5) {$t_n$};
    \node at (\xmin+1.5,\y+0.5) {\footnotesize $t_n+\dfrac{\Delta t}{m}$};
    \node at (\xmax-1.5,\y+0.5) {\footnotesize $t_{n+1}-\dfrac{\Delta t}{m}$};
    \node at (\xmax,\y-0.5) {$t_{n+1}$};
    \foreach \y in {0,-4,-10,-12,-14,-16} \draw (-1,\y) -- (0,\y); 
    \foreach \y in {0,-4,-10,-12,-14,-16} \draw (0,\y) -- (2.5,\y);
    \foreach \y in {0,-4,-10,-12,-14,-16} \node at (3,\y) {$\dots$};
    \foreach \y in {0,-4,-10,-12,-14,-16} \draw (3.5,\y) -- (6,\y);
    \foreach \y in {0,-4,-10,-12,-14,-16} \draw[->] (6,\y) -- (7,\y); 
    \foreach \y in {0,-4,-10,-12,-14,-16} \node[right] at (7,\y) {$t$};
    \def \ymin {-16}
    \def \ymax {0}
    \def \xmin {0}
    \def \xmax {6}
    \draw[dotted, color=gray] (\xmin    ,\ymin) -- (\xmin    ,\ymax);
    \draw[dotted, color=gray] (\xmin+1.5,\ymin) -- (\xmin+1.5,\ymax);
    \draw[dotted, color=gray] (\xmax-1.5,\ymin) -- (\xmax-1.5,\ymax);
    \draw[dotted, color=gray] (\xmax    ,\ymin) -- (\xmax    ,\ymax);
    \def \y {-16}
    \draw[->] (-3.8,0) -- (-3.8,\y);
    \node[rotate=90] at (-3.8-0.5,\y/2) {PIC loop operations for hybrid PIC-JR$_{ho}m$ algorithm within $[t_n,t_{n+1}]$};
\end{tikzpicture}
\caption{Diagram of the alternating nodal-staggered PIC-JR$_{ho}m$ algorithm.}
\label{fig:pic_loop_hybrid}
\end{figure}

Just like the standard and averaged formulations of PSATD PIC, the PIC-JR$_{ho}$m algorithm can readily adopt the ``hybrid nodal-staggered'' scheme presented in~\cite{ZoniCPC2022} where the field alternate between nodal and staggered representations on the simulation grid. More precisely, the Maxwell solve and guard cell exchanges are performed on a staggered ``Yee'' grid while the charge/current depositions and fields gather are performed with field quantities on a separate nodal grid. This ``hybrid'' alternating nodal-staggered extension allows to retain the advantages of low numerical dispersion and compact stencils of the integration of Maxwell's equations on a  staggered grid with the stability associated with the interpolation of fields onto the particles from a nodal grid~\cite{ZoniCPC2022} (esp. for NCI-prone boosted-frame simulations). 
The application of the ``hybrid'' alternating nodal-staggered scheme to PIC-JR$_{ho}$m leads to the steps shown in Figure~\ref{fig:pic_loop_hybrid}.

\subsubsection{
\label{subsec:psatd_averaged_equations_derivation}
Extension to the time-averaged PSATD PIC algorithm}

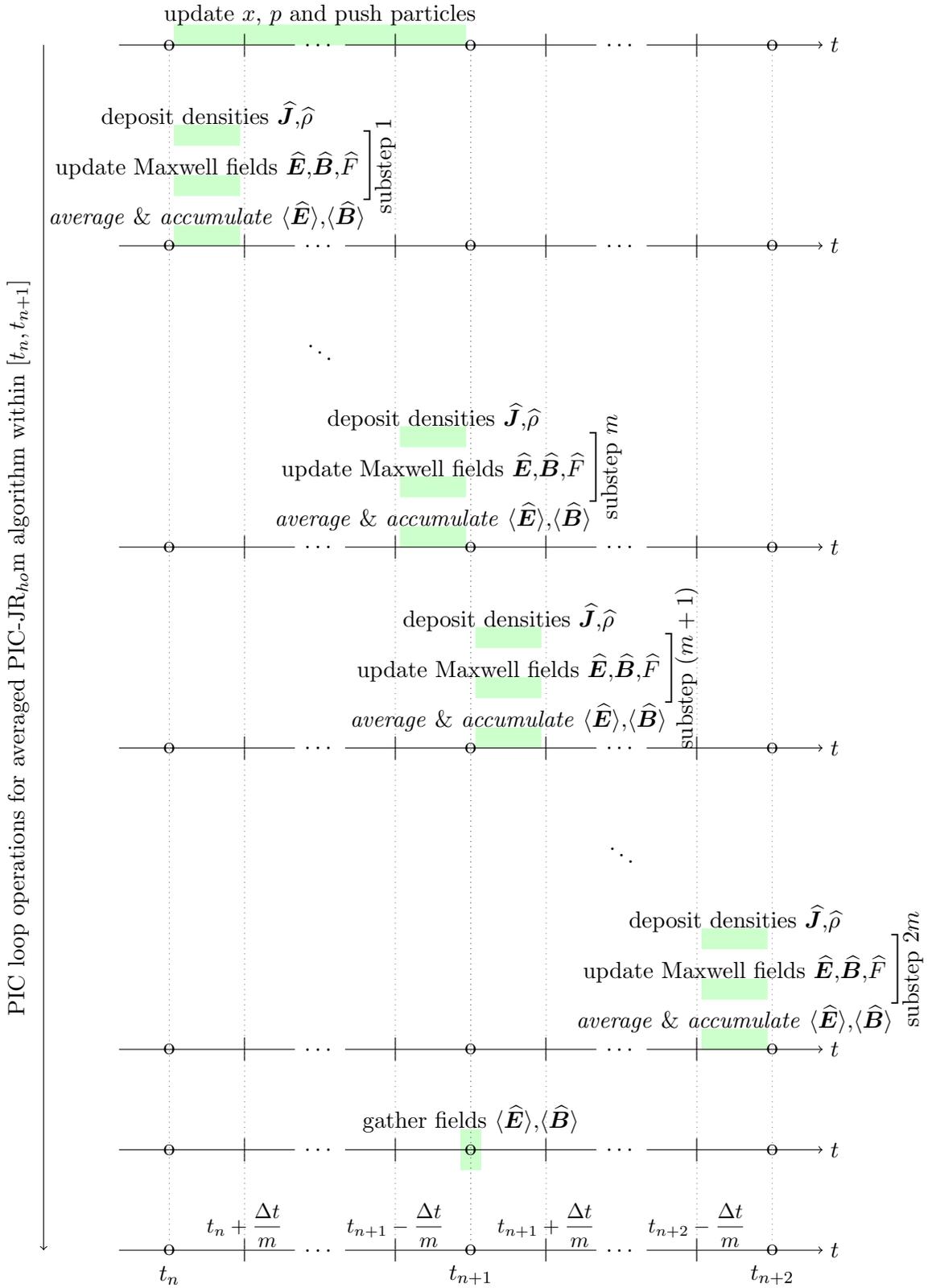
\begin{figure}[htb!]
\centering
\begin{tikzpicture}[scale=0.83]
    \def \y {0}
    \filldraw[fill=green!20, draw=green!20] (0.1,\y) rectangle (6-0.1,\y+0.4);
    \node at (3,\y+0.6) {update $x$, $p$ and push particles};
    \def \xmin {0}
    \def \xmax {1.5}
    \def \xmid {\xmin + \xmax/2 - \xmin/2}
    \def \y {-2}
    \filldraw[fill=green!20, draw=green!20] (\xmin+0.1,\y) rectangle (\xmax-0.1,\y+0.4);
    \node at (\xmid,\y+0.6) {deposit densities $\wh \Jb$,$\wh \rho$};
    \def \y {-3}
    \filldraw[fill=green!20, draw=green!20] (\xmin+0.1,\y) rectangle (\xmax-0.1,\y+0.4);
    \node at (\xmid,\y+0.6) {update Maxwell fields $\wh \Eb$,$\wh \Bb$,$\wh F$};
    \def \y {-4}
    \filldraw[fill=green!20, draw=green!20] (\xmin+0.1,\y) rectangle (\xmax-0.1,\y+0.4);
    \node at (\xmid,\y+0.6) {\textit{average} \& \textit{accumulate} $\langle\wh \Eb\rangle$,$\langle \wh \Bb\rangle$};
    \node at (4,-2.4) {$\Bigg]$};
    \node[rotate=90] at (4+0.3,-2.4) {substep $1$};
    \node at (3,-6) {$\ddots$};
    \def \xmin {4.5}
    \def \xmax {6}
    \def \xmid {\xmin + \xmax/2 - \xmin/2}
    \def \y {-8}
    \filldraw[fill=green!20, draw=green!20] (\xmin+0.1,\y) rectangle (\xmax-0.1,\y+0.4);
    \node at (\xmid,\y+0.6) {deposit densities $\wh \Jb$,$\wh \rho$};
    \def \y {-9}
    \filldraw[fill=green!20, draw=green!20] (\xmin+0.1,\y) rectangle (\xmax-0.1,\y+0.4);
    \node at (\xmid,\y+0.6) {update Maxwell fields $\wh \Eb$,$\wh \Bb$,$\wh F$};
    \def \y {-10}
    \filldraw[fill=green!20, draw=green!20] (\xmin+0.1,\y) rectangle (\xmax-0.1,\y+0.4);
    \node at (\xmid,\y+0.6) {\textit{average} \& \textit{accumulate} $\langle \wh \Eb\rangle$,$\langle\wh \Bb\rangle$};
    \node at (8.5,-8.4) {$\Bigg]$};
    \node[rotate=90] at (8.5+0.3,-8.4) {substep $m$};
    \def \xmin {6}
    \def \xmax {7.5}
    \def \xmid {\xmin + \xmax/2 - \xmin/2}
    \def \y {-12}
    \filldraw[fill=green!20, draw=green!20] (\xmin+0.1,\y) rectangle (\xmax-0.1,\y+0.4);
    \node at (\xmid,\y+0.6) {deposit densities $\wh \Jb$,$\wh \rho$};
    \def \y {-13}
    \filldraw[fill=green!20, draw=green!20] (\xmin+0.1,\y) rectangle (\xmax-0.1,\y+0.4);
    \node at (\xmid,\y+0.6) {update Maxwell fields $\wh \Eb$,$\wh \Bb$,$\wh F$};
    \def \y {-14}
    \filldraw[fill=green!20, draw=green!20] (\xmin+0.1,\y) rectangle (\xmax-0.1,\y+0.4);
    \node at (\xmid,\y+0.6) {\textit{average} \& \textit{accumulate} $\langle\wh \Eb\rangle$,$\langle\wh \Bb\rangle$};
    \node at (10,-12.4) {$\Bigg]$};
    \node[rotate=90] at (10+0.3,-12.4) {substep $(m+1)$};
    \node at (9,-16) {$\ddots$};
    \def \xmin {10.5}
    \def \xmax {12}
    \def \xmid {\xmin + \xmax/2 - \xmin/2}
    \def \y {-18}
    \filldraw[fill=green!20, draw=green!20] (\xmin+0.1,\y) rectangle (\xmax-0.1,\y+0.4);
    \node at (\xmid,\y+0.6) {deposit densities $\wh \Jb$,$\wh \rho$};
    \def \y {-19}
    \filldraw[fill=green!20, draw=green!20] (\xmin+0.1,\y) rectangle (\xmax-0.1,\y+0.4);
    \node at (\xmid,\y+0.6) {update Maxwell fields $\wh \Eb$,$\wh \Bb$,$\wh F$};
    \def \y {-20}
    \filldraw[fill=green!20, draw=green!20] (\xmin+0.1,\y) rectangle (\xmax-0.1,\y+0.4);
    \node at (\xmid,\y+0.6) {\textit{average} \& \textit{accumulate} $\langle\wh \Eb\rangle$,$\langle\wh \Bb\rangle$};
    \node at (14.5,-18.4) {$\Bigg]$};
    \node[rotate=90] at (14.5+0.3,-18.4) {substep $2m$};
    \def \xmid {6}
    \def \y {-22}
    \filldraw[fill=green!20, draw=green!20] (\xmid-0.2,\y-0.4) rectangle (\xmid+0.2,\y+0.4);
    \node at (\xmid,\y+0.6) {gather fields $\langle\wh \Eb\rangle$,$\langle\wh \Bb\rangle$};
    \def \xmin {0}
    \def \xmax {6}
    \foreach \y in {0,-4,-10,-14,-20,-22,-24} \node at (\xmin,\y) {o};
    \foreach \y in {0,-4,-10,-14,-20,-22,-24} \node at (\xmin+1.5,\y) {$\vert$};
    \foreach \y in {0,-4,-10,-14,-20,-22,-24} \node at (\xmax-1.5,\y) {$\vert$};
    \foreach \y in {0,-4,-10,-14,-20,-22,-24} \node at (\xmax,\y) {o};
    \def \y {-24}
    \node at (\xmin,\y-0.5) {$t_n$};
    \node at (\xmin+1.5,\y+0.5) {\footnotesize $t_n+\dfrac{\Delta t}{m}$};
    \node at (\xmax-1.5,\y+0.5) {\footnotesize $t_{n+1}-\dfrac{\Delta t}{m}$};
    \node at (\xmax,\y-0.5) {$t_{n+1}$};
    \def \xmin {6}
    \def \xmax {12}
    \foreach \y in {0,-4,-10,-14,-20,-22,-24} \node at (\xmin+1.5,\y) {$\vert$};
    \foreach \y in {0,-4,-10,-14,-20,-22,-24} \node at (\xmax-1.5,\y) {$\vert$};
    \foreach \y in {0,-4,-10,-14,-20,-22,-24} \node at (\xmax,\y) {o};
    \def \y {-24}
    \node at (\xmin+1.5,\y+0.5) {\footnotesize $t_{n+1}+\dfrac{\Delta t}{m}$};
    \node at (\xmax-1.5,\y+0.5) {\footnotesize $t_{n+2}-\dfrac{\Delta t}{m}$};
    \node at (\xmax,\y-0.5) {$t_{n+2}$};
    \foreach \y in {0,-4,-10,-14,-20,-22,-24} \draw (-1,\y) -- (0,\y); 
    \foreach \y in {0,-4,-10,-14,-20,-22,-24} \draw (0,\y) -- (2.5,\y);
    \foreach \y in {0,-4,-10,-14,-20,-22,-24} \node at (3,\y) {$\dots$};
    \foreach \y in {0,-4,-10,-14,-20,-22,-24} \draw (3.5,\y) -- (6,\y);
    \foreach \y in {0,-4,-10,-14,-20,-22,-24} \draw (6,\y) -- (8.5,\y);
    \foreach \y in {0,-4,-10,-14,-20,-22,-24} \node at (9,\y) {$\dots$};
    \foreach \y in {0,-4,-10,-14,-20,-22,-24} \draw (9.5,\y) -- (12,\y);
    \foreach \y in {0,-4,-10,-14,-20,-22,-24} \draw[->] (12,\y) -- (13,\y); 
    \foreach \y in {0,-4,-10,-14,-20,-22,-24} \node[right] at (13,\y) {$t$};
    \def \ymin {-24}
    \def \ymax {0}
    \def \xmin {0}
    \def \xmax {6}
    \draw[dotted, color=gray] (\xmin    ,\ymin) -- (\xmin    ,\ymax);
    \draw[dotted, color=gray] (\xmin+1.5,\ymin) -- (\xmin+1.5,\ymax);
    \draw[dotted, color=gray] (\xmax-1.5,\ymin) -- (\xmax-1.5,\ymax);
    \draw[dotted, color=gray] (\xmax    ,\ymin) -- (\xmax    ,\ymax);
    \def \xmin {6}
    \def \xmax {12}
    \draw[dotted, color=gray] (\xmin    ,\ymin) -- (\xmin    ,\ymax);
    \draw[dotted, color=gray] (\xmin+1.5,\ymin) -- (\xmin+1.5,\ymax);
    \draw[dotted, color=gray] (\xmax-1.5,\ymin) -- (\xmax-1.5,\ymax);
    \draw[dotted, color=gray] (\xmax    ,\ymin) -- (\xmax    ,\ymax);
    \def \y {-24}
    \draw[->] (-2.5,0) -- (-2.5,\y);
    \node[rotate=90] at (-2.5-0.5,\y/2) {PIC loop operations for averaged PIC-JR$_{ho}$m algorithm within $[t_n,t_{n+1}]$};
\end{tikzpicture}
\caption{Diagram of the time-averaged PIC-JR$_{ho}m$ algorithm.}
\label{fig:pic_loop_averaged}
\end{figure}

In Refs.~\cite{ShapovalPRE2021}, an extension to PSATD PIC, named time-averaged PSATD PIC (also labeled as averaged PIC for convenience), is presented that enables stable boosted-frame simulations even when the time step is larger than the Courant condition along a given axis, e.g., $c\Delta t=\Delta z > \Delta x$.
With the time-averaged algorithm, the field quantities that are gathered onto the particles are given by time averages of the fields on the grid obtained by analytically integrating the $\wh \Eb$ and $ \wh \Bb$ fields from $t=n\Delta t$ to $t=(n+2)\Delta t$. The time-averaged PIC-JR$_{ho}$m algorithm consists of the steps shown in Figure~\ref{fig:pic_loop_averaged}, where 
%
%
the analytical average of $\wh \Eb$ and $\wh \Bb$ at time $t=(n+1)\Delta t$ are,
\begin{subequations}
\begin{align}
\langle \wh \Eb^{n+1} \rangle &= \frac{1}{2\Delta t} \sum_{\ell=0}^{2m-1}  \Big[ \frac{S}{ ck } \wh{\Eb}^{n+\ell/m} +  \frac{i c^2 Y_4}{c^2k^2} \kb \times \wh \Bb^{n+\ell/m}  + \frac{i Y_4}{2ck\delta t} \wh{F}^{n+\ell/m}  \nonumber  \\
& + \frac{1}{\varepsilon_0c^2k^2}(  Y_1 \ab_{J}^{\tau} - Y_5  \bb_J^{\tau}  - Y_4  \cb_J^{\tau} )  - i c^2 \kb \times (  Y_6 a_{\rho}^{\tau} + Y_7 b_{\rho}^{\tau}  + Y_8  c_{\rho}^{\tau} )\Big] \,, \\
\langle \wh \Bb^{n+1} \rangle &= \frac{1}{2\Delta t} \sum_{\ell=0}^{2m-1}  \Big[ \frac{S}{ck}\wh{\Bb}^{n+\ell/m} - \frac{i Y_4}{c^2k^2} \kb \times \wh \Eb^{n+\ell/m}  + i \kb \times (  Y_6 \ab_J^{\tau} + Y_7 \bb_J^{\tau}  + Y_8  \cb_J^{\tau} )\Big] \,.
\end{align}
\label{EB_avg}
\end{subequations}
For a detailed derivation see Appendix~\ref{sec:Appendix_averaged_eqs}.
%
%

\subsection{
\label{subsec:relation_to_galilean}
Relation to the Galilean PSATD PIC algorithm}
This section examines the relationship between the Galilean PIC algorithm, the standard PSATD PIC algorithm and the new PIC-JR$_{ho}$m algorithm. To this end, it is instructive to ``deconstruct'' the Galilean PIC algorithm by separating it in two independent steps: (i) a shift of the quantities to recenter them on a grid moving at $\vbgal$, (ii) the integration of the PSATD equations assuming that the current source is constant along the flow moving at the Galilean velocity $\vbgal$.

The standard Galilean PIC scheme~\cite{LehePRE2016,KirchenPoP2016} can then be written highlighting terms that arise from step (i) in red in Eqs.(~\ref{eq_update_galilean_B})-(\ref{eq_update_galilean_E}) and those from step (ii) in blue in Eqs.(~\ref{eq_X1})-(\ref{eq_X4}) and Eq.(~\ref{eq_chi1}): 
\begin{subequations}
\begin{align}
\label{eq_update_galilean_B}
\wh\Bb^{n+1} & = C {\highlightred{\theta^2}} \wh\Bb^{n}
- i \, \frac{S}{\omega} \, \kb \times (\highlightred{ \theta^2} \wh\Eb^{n})
+ i \, X_1 \, \kb \times \highlightred{ \theta}\wh\Jb^{n+\frac 12} \,, \\[5pt]
\label{eq_update_galilean_E}
\wh\Eb^{n+1} & = C \highlightred{ \theta^2} \wh\Eb^{n}
+ i \, c^2 \, \frac{S}{\omega} \, \kb \times (\highlightred{ \theta^2} \wh\Bb^{n})
+ X_4 \, \highlightred{ \theta}\wh\Jb^{n+\frac 12}
+ i \, \left( X_3 \, \highlightred{ \theta^2} \wh\rho^{\, n} - X_2 \, \wh\rho^{\, n+1}\right) \, \kb \,,
%
\end{align}
\end{subequations}
where the coefficients $X_1$, $X_2$, $X_3$ and $X_4$ are defined as
\begin{subequations}
\begin{align}
\label{eq_X1}
& X_1 := \frac{1}{\eps_0 (\omega^2 \highlightblue{- \Omega^2} )} \left(\highlightblue{\theta^*} - \highlightblue{\theta} \, C \highlightblue{+ i \, \Omega \, \theta \, \frac{S}{\omega}} \right) \,, \\[5pt]
\label{eq_X2}
& X_2 := \frac{c^2}{\highlightblue{\theta^* - \theta}}
\left(\highlightblue{\theta^*} \frac{\, \chi_1}{\eps_0 \, \omega^2} \highlightblue{- \theta \frac{1 - C}{\eps_0 \, \omega^2}}\right) \,, \\[5pt]
\label{eq_X3}
& X_3 := \frac{c^2}{\highlightblue{\theta^* - \theta}}
\left(\highlightblue{\theta^*} \frac{\, \chi_1}{\eps_0 \, \omega^2} \highlightblue{- \theta^* \frac{1 - C}{\eps_0 \, \omega^2}}\right) \,, \\[5pt]
\label{eq_X4}
& X_4 := \highlightblue{i \, \Omega \, X_1} - \frac{\highlightblue{\theta}}{\eps_0} \frac{S}{\omega} \,,
%
%
\end{align}
\end{subequations}
with
\begin{equation}
\label{eq_chi1}
\chi_1 := \frac{\highlightblue{\omega^2}}{\omega^2 \highlightblue{- \Omega^2}} \left(\highlightblue{\theta^* - \theta \, C} + \highlightblue{i \, \Omega \, \theta} \, \frac{S}{\omega} \right) \,,
%
%
\end{equation}
where $\Omega := \vbgal \cdot \kb$, $\omega := c \, k$, 
$C := \cos(\omega \, \dt)$,
$S := \sin(\omega \, \dt)$,
$\theta := e^{i \Omega \dt / 2}$ and
$\theta^* := e^{- i \Omega \dt / 2}$.

When setting $\vbgal=0$, the system (\ref{eq_update_galilean_B})-(\ref{eq_chi1}) converges to the standard PSATD algorithm, as expected. 

Step (i), which corresponds to the multiplication of some of the terms by $\theta$ or $\theta^2$, in red in (\ref{eq_update_galilean_B})-(\ref{eq_update_galilean_E}), is the easiest to interpret: noting that a multiplication by $\theta := e^{i \Omega \dt / 2}$ in Fourier space corresponds to shifting the terms spatially by the distance $\vbgal \dt/2$ in real space, the terms known at time $n+1/2$ are multiplied by $\theta$, hence shifted by $\vbgal \dt/2$ while the terms known at time $n$ are multiplied by $\theta^2$, hence shifted by $\vbgal \dt$. These are exactly the shifts that are needed to bring the corresponding quantities to their new grid location after one time step, when assuming a Galilean frame of reference moving at $\vbgal$.

Understanding the terms associated with step (ii) requires a more detailed comparisons between how the standard and the Galilean PIC equations are obtained. The standard PSATD algorithm is derived assuming that the current density (source term) is constant over one time step on a fixed grid. The Galilean algorithm makes the same assumption but in a Galilean frame, i.e., that the current density (source term) is constant over one time step on a Galilean grid. Following this comparison, it flows logically that step (ii) ought to correspond to an integration of the PSATD equations {\bf on a fixed grid} assuming that the currents are constant {\bf along a segment of length $\vbgal \Delta t$}. Indeed, it was verified that integrating the PSATD equations based on these assumptions leads to the system (\ref{eq_update_galilean_B})-(\ref{eq_chi1}) with the terms highlighted in red replaced by $1$ in (\ref{eq_update_galilean_B})-(\ref{eq_update_galilean_E}).

From this, it follows that the new algorithm PIC-JR$_{ho}$m is related to step (ii) of the Galilean PIC algorithm in the following way. While 
step (ii) of Galilean PIC provides a more accurate analytical integration of the PSATD equations over one time step for a flow that moves uniformly at $v_{gal}$, the PIC-JR$_{ho}$m, with its arbitrary time-dependence of $\Jb$ and $\rho$ and its subintervals, provides a more accurate analytical integration of the PSATD equations over one time step for an arbitrary local flow of particles. The new PIC-JR$_{ho}$m algorithm can thus be viewed as a possible generalization of step (ii) of the Galilean PIC algorithm. Indeed, the numerical tests discussed below show that, like the Galilean PIC algorithm, PIC-JR$_{ho}$m can lead to simulations that are very stable with regard to the numerical Cherenkov instability, and that it can also remain accurate in cases where the Galilean assumption is becoming less appropriate.

\section{Numerical tests}
\label{sec:numerical_tests}
This section presents various physics applications to test the novel PIC-JR$_{ho}$m algorithm.
All simulations and results have been performed and obtained with the open-source electromagnetic PIC code WarpX \citep{Vay2018,Vay2020,Vay2021}.
The current implementation provides the flexibility to:
\begin{itemize}
\item choose an arbitrary polynomial time dependency of $\Jb$ and $\rho$ among the following combinations:
    \begin{itemize}
        \item $\Jb$ and $\rho$ constant in time (CC$m$);
        \item $\Jb$ constant in time and $\rho$ linear in time (CL$m$);
        \item $\Jb$ and $\rho$ linear in time (LL$m$);
        \item $\Jb$ and $\rho$ quadratic in time (QQ$m$);
    \end{itemize}
    \item choose the number of subintervals $m$ within one time step;
    \item turn on/off the divergence cleaning term, that is, solve Maxwell's equations \eqref{Maxwell} with or without the scalar field $F$;
    \item turn on/off the time averaging of the $\Eb$ and $\Bb$ fields gathered on the macro-particles, as in \eqref{EB_avg}.
\end{itemize}

To assess the stability of the novel PIC-JR$_{ho}$m method theoretically, the analytical dispersion equation was derived (see appendix).
This allows to predict the growth rates of the numerical Cherenkov instability in the case of a uniform drifting plasma.
Moreover, a variety of WarpX simulation tests were run to further investigate the method's stability and accuracy.
These tests include: 2D simulations of a uniform plasma drifting with a relativistic velocity $\vb_0$ (with/without divergence cleaning, with/without subintervals, and with small/large time steps) and 3D simulations of laser wakefield acceleration (LWFA).
\subsection{Stability of a uniform plasma drifting at relativistic velocity}
\begin{figure}
\includegraphics[width=\linewidth]{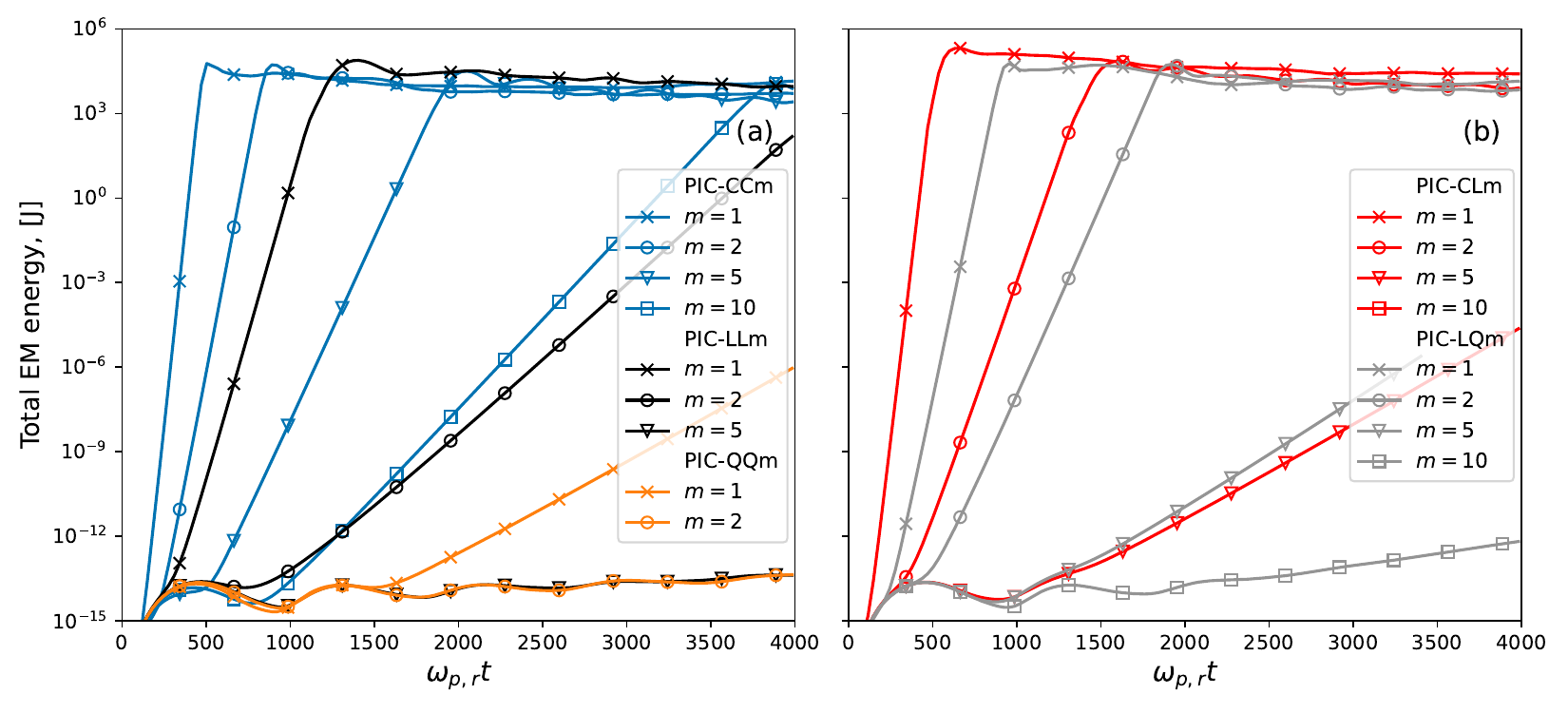}
\caption{\textbf{WarpX simulations of a uniform plasma with a time step at the Courant condition limit $c\Delta t = \Delta x = \Delta z$ and a stencil at infinite order, without divergence cleaning}. The total electromagnetic (EM) energy of a uniform plasma drifting at relativistic velocity $\vb_0$ along the $z$-axis is plotted versus the time of the simulation with (a) the same time-dependencies for $\Jb$ and $\rho$ and (b) different time-dependencies for $\Jb$ and $\rho$, 
for various combinations of time-dependency and number of timestep subintervals.
Here, $\omega_{p,r} = \omega_p /\sqrt{\gamma_0}$ is the relativistic plasma frequency, 
where time $\omega_{p,r}t=4000$ corresponds to roughly $6.7\times 10^4$ time steps. 
}
\label{fig:unif_pl_small_dt}
\end{figure}

\begin{figure}
\includegraphics[width=\linewidth]{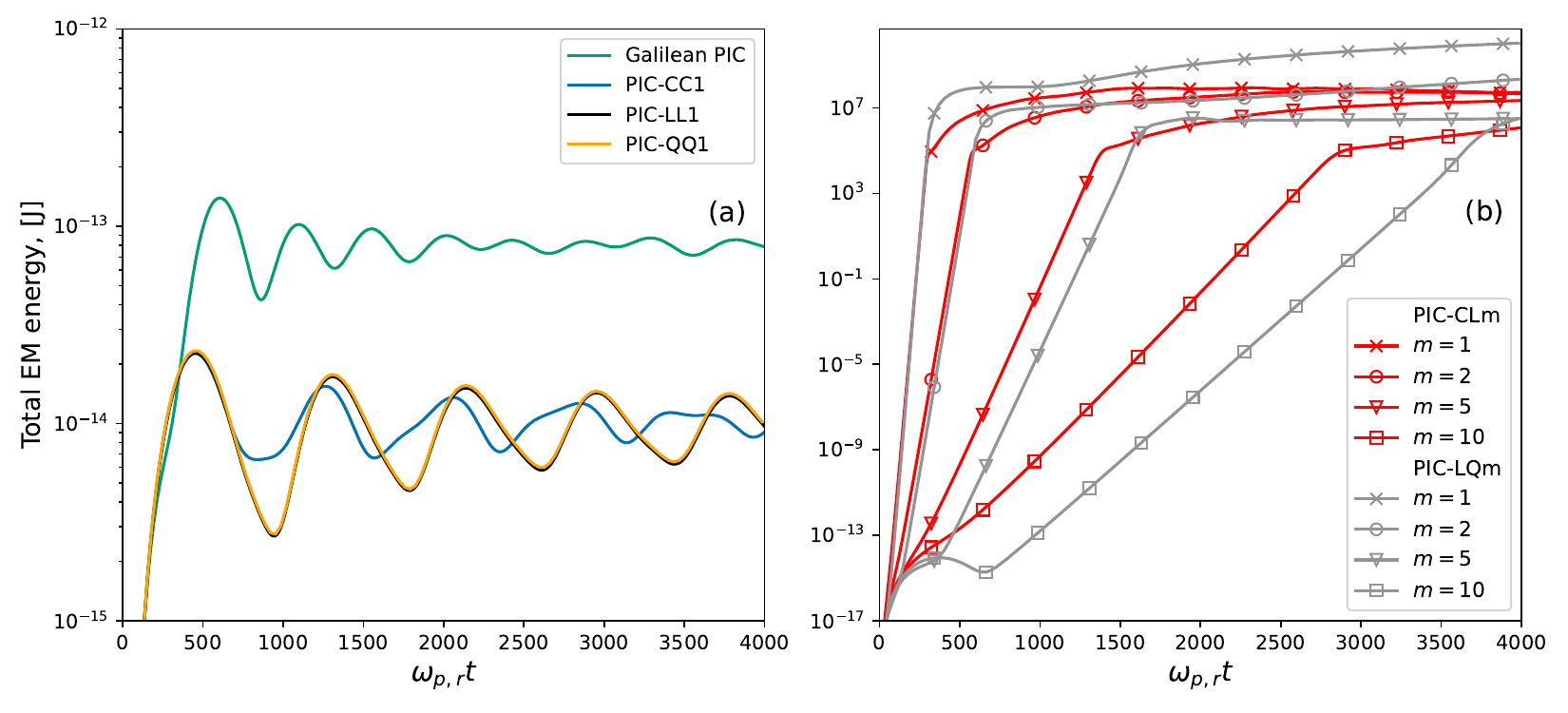}
\caption{\textbf{WarpX simulations of a uniform plasma with a time step at the Courant limit $c\Delta t = \Delta x = \Delta z$ and a stencil at infinite order, with divergence cleaning}. The total electromagnetic (EM) energy of a uniform plasma drifting at relativistic velocity $\vb_0$ along the $z$-axis is plotted versus the time of the simulation with (a) same time-dependencies for $\wh \Jb$ and $\wh \rho$ and (b) different time-dependencies for $\wh \Jb$ and $\wh \rho$, 
for various combinations of time-dependency and number of timestep subintervals.
The energy history from a simulation using the Galilean PIC algorithm~\cite{LehePRE2016} is also plotted for comparison in plot (a).}
\label{fig:unif_pl_small_dt_divE}
\end{figure}

This section presents WarpX simulations of a uniform electron-proton plasma with density $n_0 = \epsilon_0 m_e c^2 \gamma_0 / e^2$ (where $\epsilon_0$ is the permittivity of free space, $c$ is the speed of light in free space, and $e$ and $m_e$ are respectively the electron charge and mass), drifting along $z$ with a relativistic velocity $\vb_0 = (0,0,v_0)$, with $v_0 = c \sqrt{1 - 1/\gamma_0^2}$ and Lorentz factor $\gamma_0 = 130$, through a two-dimensional computational domain with ${x_{\min} = z_{\min} = -6.45 \, \mu\text{m}}$ and ${x_{\max} = z_{\max} = 6.45 \, \mu\text{m}}$, periodic boundary conditions and $600 \times 200$ grid cells along $x$ and $z$, respectively.
The simulations were performed with 4 particles per cell, per species, 1 pass of bilinear filter in the transverse direction $x$ and 4 passes in the longitudinal direction $z$ (the direction along which the plasma is drifting).
Four cases were considered: 
\begin{enumerate}
    \item PIC-JR$_{ho}$m with $c\Delta t=\Delta x=\Delta z$ {\bf without} divergence cleaning (Figure~\ref{fig:unif_pl_small_dt});
    \item PIC-JR$_{ho}$m with $c\Delta t=\Delta x=\Delta z$ {\bf with} divergence cleaning (Figures~\ref{fig:unif_pl_small_dt_divE} and \ref{fig:unif_pl_small_dt_theory});
    \item {\bf averaged} PIC-JR$_{ho}$m with $c\Delta t=6\Delta x=\Delta z$ with divergence cleaning (Figure~\ref{fig:unig_pl_large_dt_inifiniteorder});
    \item PIC-JR$_{ho}$m with $c\Delta t=\Delta x=\Delta z$ and averaged PIC-JR$_{ho}$m with $c\Delta t=6\Delta x=\Delta z$, with divergence cleaning and {\bf finite order stencils} (Figure~\ref{fig:unif_plasma_finite_order_small} and \ref{fig:nci_gr_finite_order});
\end{enumerate}
and are discussed below in detail.
%
%

\begin{enumerate}
\item{\bf{PIC-JR$_{ho}$m with $c\Delta t=\Delta x=\Delta z$ {\bf without} divergence cleaning}}

\begin{figure}
\centering
\begin{subfigure}[b]{\textwidth}
\centering
\includegraphics[width=\linewidth]{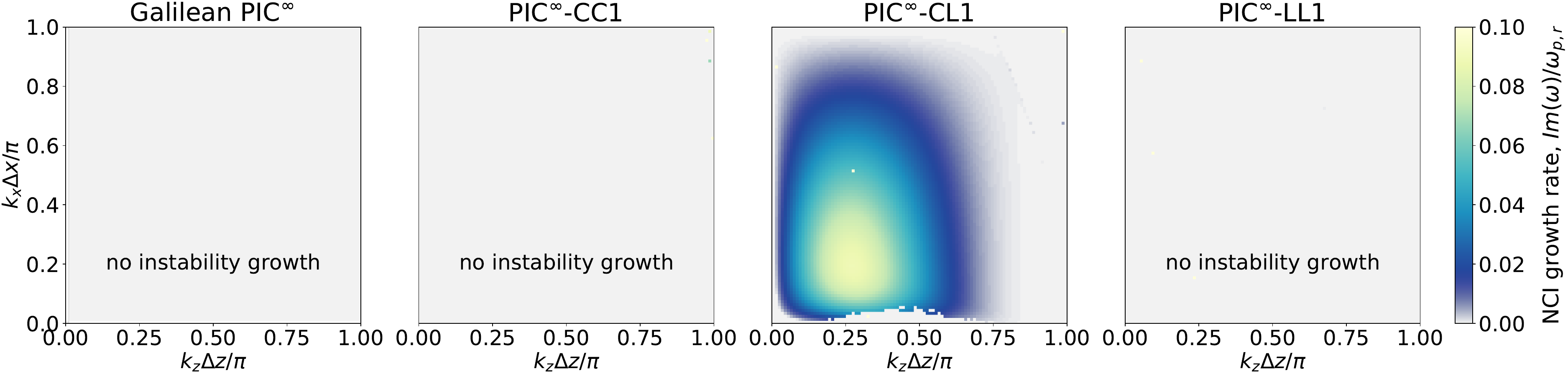}
\caption{Theory}
\end{subfigure}
\begin{subfigure}[b]{\textwidth}
\centering
\includegraphics[width=\linewidth]{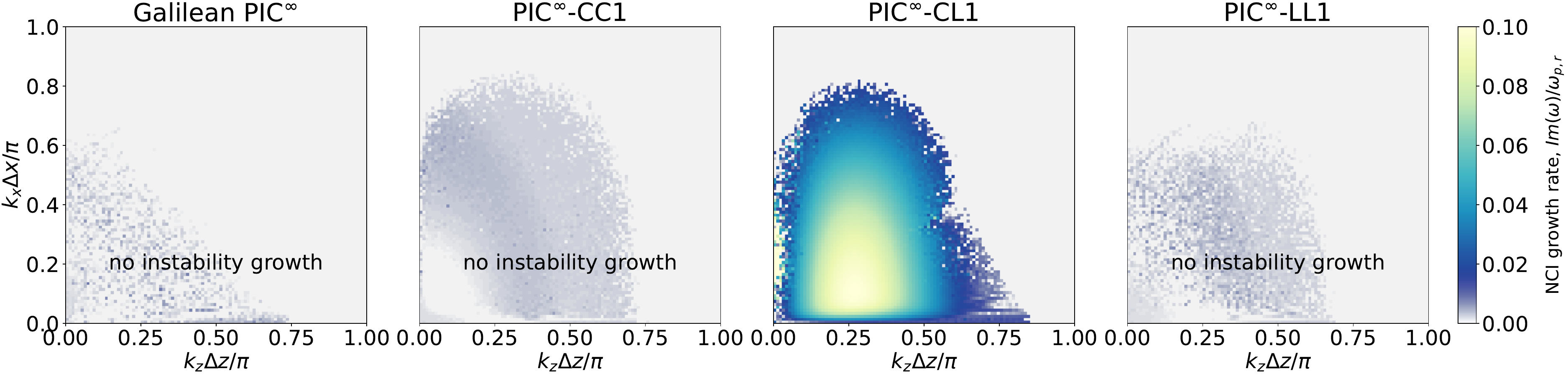}
\caption{WarpX simulations}
\end{subfigure}
\caption{\textbf{NCI growth rates, with small time step $c\Delta t = \Delta x$}. Normalized NCI growth rates  $\text{Im}(\omega)/\omega_{p,r}$ in spectral space $(k_x,k_z)$, calculated from (a) the analytical dispersion equation and (b) WarpX simulations (b) using four different solvers: Galilean PIC with matched velocity ($\vb_{gal} = \vb_0$), standard PIC-CL1,  novel PIC-CC1 and PIC-LL1. All numerical and physical parameters are the same as in Figure~\ref{fig:unif_pl_small_dt_divE}: divergence cleaning is used in all cases except for Galilean PIC. The simulation time step is $c\Delta t =\Delta x =\Delta z$ and the transverse and longitudinal cell sizes are $\Delta x = 6.45 \times 10^{-2} \, k_{p,r}^{-1}$, where $k_{p,r}^2 = n_0 e^2/(\epsilon_0 m_e c^2 \gamma_0)$.}
\label{fig:unif_pl_small_dt_theory}
\end{figure}
Figure~\ref{fig:unif_pl_small_dt} shows the total electromagnetic energy as a function of $\omega_{p,r} t = \omega_p t /\sqrt{\gamma_0} = \sqrt{e^2n_0/(m_e\epsilon_0)}$ 
obtained from WarpX simulations using PIC-JR$_{ho}$m with CC$m$, LL$m$, QQ$m$, CL$m$ and LQ$m$, for $m=1,2,5,10$, 
without divergence cleaning. 
In this case, 
increasing the order of the polynomial dependency (from C, L to Q), or timestep subintervals ($m>1$), helps delaying the onset of the instability and lowering the growth rate. 
When using the same time dependency for $\Jb$ and $\rho$ (CC, LL and QQ), for a given number of depositions per step, it is more advantageous to increase the order of the polynomial than to increase the subintervals number $m$. Conversely, when using a different time dependency for $\Jb$ and $\rho$ (CL, LQ), it is more advantageous to increase the number of subintervals $m$ than to increase the order of the polynomial. 
Matching the time dependency of $\wh \Jb$ and $\wh \rho$ (as in CC, LL, QQ) is also increasing stability, with PIC-LL5 and PIC-QQ2 being more stable than PIC-LQ10.

\item{\bf PIC-JR$_{ho}$m with $c\Delta t=\Delta x=\Delta z$ with divergence cleaning}

Figure~\ref{fig:unif_pl_small_dt_divE} shows the total electromagnetic energy as a function of $\omega_{p,r} t$ obtained from WarpX simulations using PIC-JR$_{ho}$m with CC$m$, LL$m$, QQ$m$, CL$m$ and LQ$m$ for $m=1,2,5,10$, with divergence cleaning. 
The energy history from a simulation using the Galilean PIC algorithm~\cite{LehePRE2016} is also plotted for comparison. 

In contrast to the previous case, when divergence cleaning is used, having the same time dependency for $\wh \Jb$ and $\wh \rho$ leads to an extraordinary level of stability  
that is comparable to the one of the Galilean PSATD method.
Conversely, turning on the divergence cleaning degrades significantly the stability when using different time dependencies for $\wh \Jb$ and $\wh \rho$ (CL and LQ). 

The remarkable stability reported in Figure~\ref{fig:unif_pl_small_dt_divE} when matching the time-dependencies is confirmed with a theoretical NCI analysis.
Figure~\ref{fig:unif_pl_small_dt_theory} shows the NCI growth rates, $\text{Im}(\omega)/\omega_{p,r}$, obtained from theoretical calculations and WarpX simulations for the Galilean PIC, the standard PSATD PIC (CL1), PIC-CC1 and PIC-LL1, with an excellent agreement between theory and simulations.

A detailed derivation of the two-dimensional dispersion equation for the PIC-JR$_{ho}$m scheme,  for time dependencies of $\wh \Jb$ and $\wh \rho$ up to quadratic, is presented in Appendix~\ref{sec:Appendix_dispeq}, clarifying the origin of the remarkable stability that is observed with PIC-CC1, PIC-LL1 and PIC-QQ1.
As explained in the appendix, it can be shown that under some conditions that include having the same time dependency for $\wh \Jb$ and $\wh \rho$, key terms cancel out in the analysis matrix, leading to stable real solutions of the determinant. 
%

\item{\bf{Averaged PIC-JR$_{ho}$m with $c\Delta t=6\Delta x=\Delta z$ with divergence cleaning}}
\begin{figure}
\includegraphics[width=\linewidth]{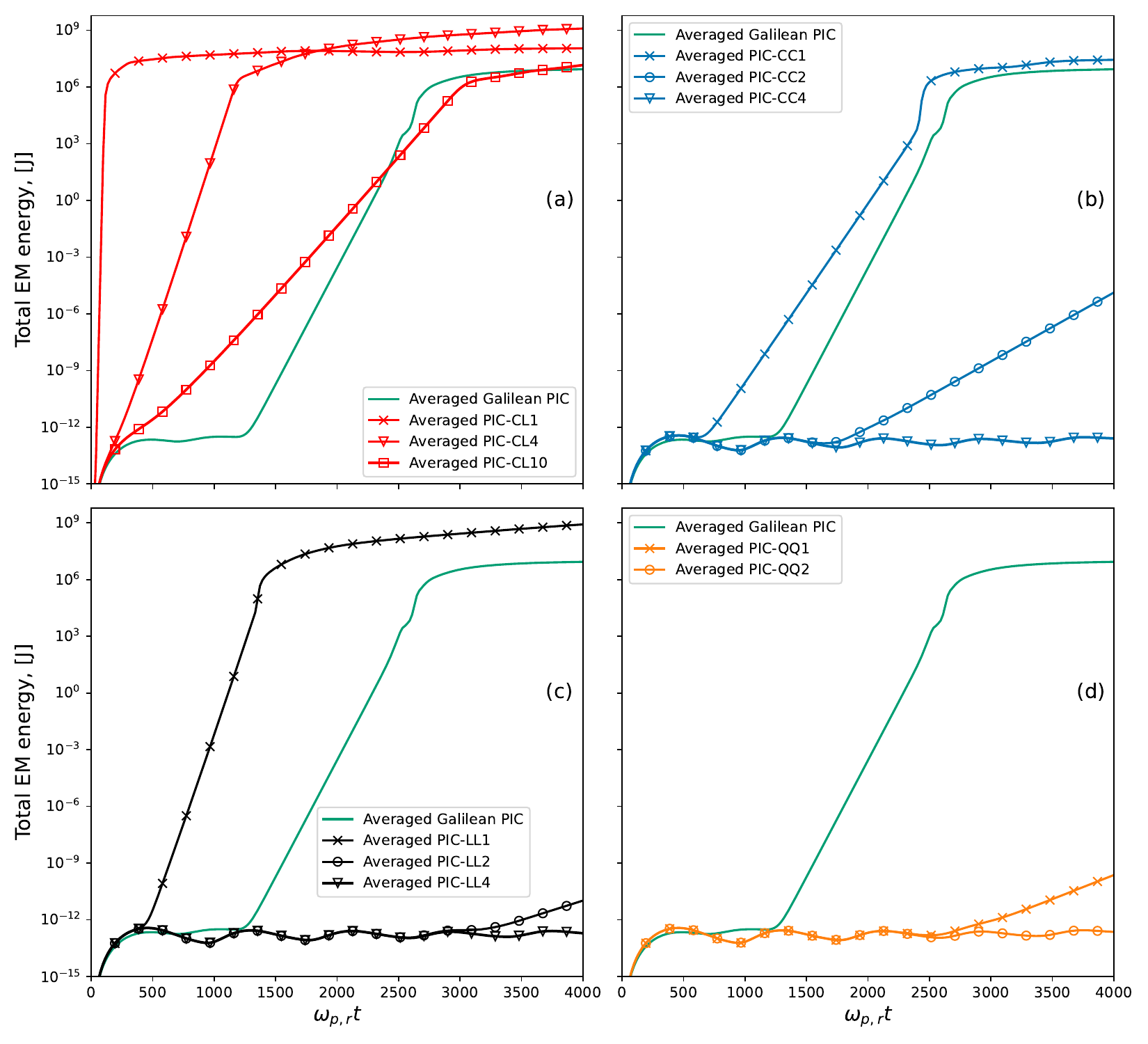}
\caption{\textbf{WarpX simulation of a uniform plasma with $c\Delta t=\Delta z=6\Delta x$ }. Total electromagnetic (EM) energy of a uniform plasma drifting at relativistic velocity $\vb_0$ along the $z$-axis. Simulations were performed with time steps of $c\Delta t = \Delta z = 6\Delta x$ and divergence cleaning, for (a) PIC-CL$m$, (b) PIC-CC$m$, (c) PIC-LL$m$ and (d) PIC-QQ$m$, with $m=1,2,4$. The results from a simulation using the average Galilean PIC solver is also plotted for comparison. 
}
\label{fig:unig_pl_large_dt_inifiniteorder}
\end{figure}
\\
In this test, the transverse cell size is intentionally set to a much smaller value than the longitudinal cell size, as typical in plasma accelerator simulations in a Lorentz boosted frame of reference~\cite{Vayprl07,Vaypop2011} with a high Lorentz factor $\gamma_{0}$~\cite{ShapovalPRE2021}, while keeping the time step at the CFL limit of the longitudinal cell size: $c\Delta t=\Delta z=6\Delta x$. The results from Figure~\ref{fig:unig_pl_large_dt_inifiniteorder} show that this case is more challenging for all schemes, and even the averaged Galilean PIC scheme is not stable beyond 1000 plasma periods. Increasing the order of the polynomial and the number of subintervals $m$ both help delaying the onset and lowering the growth rate of the instability, slowly for CL$m$ but quite effectively for CC$m$, LL$m$ and QQ$m$, with increasing the number of subintervals $m$ being the most effective strategy for a given number of depositions per time step. 

\item {\bf PIC$^p$-JR$_{ho}$m with $c\Delta t=\Delta x=\Delta z$ and averaged PIC$^p$-JR$_{ho}$m with $c\Delta t=6\Delta x=\Delta z$, with divergence cleaning and finite order stencils}\\
This test shows numerical (Figure~\ref{fig:unif_plasma_finite_order_small}) and theoretical (Figure~\ref{fig:nci_gr_finite_order}) evidence that using a stencil at finite-order $p$ with PIC$^p$-LL$m$ leads to a degradation of the stability that increases as the order $p$ decreases. 
This is because the NCI resonant modes, caused by temporal and spatial aliasing, depends on the stencil order:
\begin{equation}
\centering
[k_{x,res}^{p}] = \sqrt{ \left([k_z^p]\frac{v_0}{c} + m_z\frac{2\pi}{\Delta z}\frac{v_0}{c} - \frac{2\pi m_t}{c\Delta t} \right)^2 - [k_z^p]^2},  
\label{eq:res_modes}
\end{equation}
for any $m_z, m_t \in \mathbb{Z}$, where $m_z$ is the spatial alias index and $m_t$ is the temporal alias index \cite{KirchenPRE2020}.
As the stencil order gets lower, such resonant modes relocate to lower wavenumbers where the resonance is stronger, as can be seen on  Figure~\ref{fig:nci_gr_finite_order} that shows the theoretical NCI growth rate at different spectral orders, $p=8,16,32$.
A non-zero growth rate is observed solely along the NCI resonant mode that is caused by aliasing between the temporal $m_t=0$ and spatial $m_z=0$ modes. 
The results from \mbox{Figures~\ref{fig:unif_plasma_finite_order_small}-\ref{fig:nci_gr_finite_order}} indicate that the choice of stencil order will depend on the total duration of the simulations (as measured in plasma periods) for a given application. 
%
%
\begin{figure}
\includegraphics[width=\linewidth]{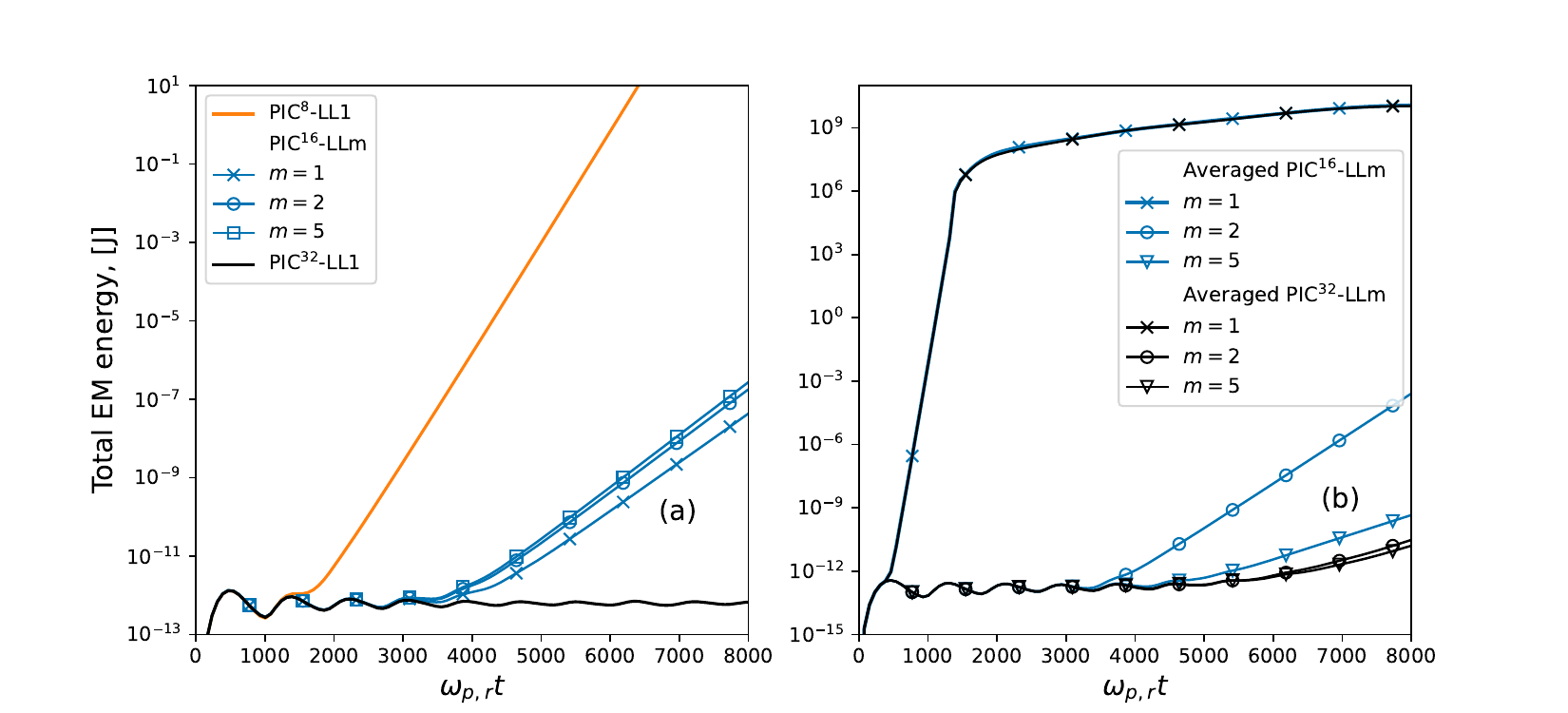}
\caption{\textbf{WarpX simulation of a uniform plasma at finite spectral order $p$}. Total electromagnetic (EM) energy of a uniform plasma drifting at relativistic velocity $\vb_0$ along the $z$-axis. Simulations were performed with (a) the standard PIC$^p$-JR$_{ho}$m algorithm with $c \Delta t = \Delta z =  \Delta x $ and (b) the averaged PIC$^p$-JR$_{ho}$m algorithm with $c \Delta t = \Delta z =  6 \Delta x$, using linear time dependency for $\wh \Jb$ and $\wh \rho$ in all cases and varying the spectral order $p=8,16,32$.}
\label{fig:unif_plasma_finite_order_small}
\end{figure}
\begin{figure}
\includegraphics[width=\linewidth]{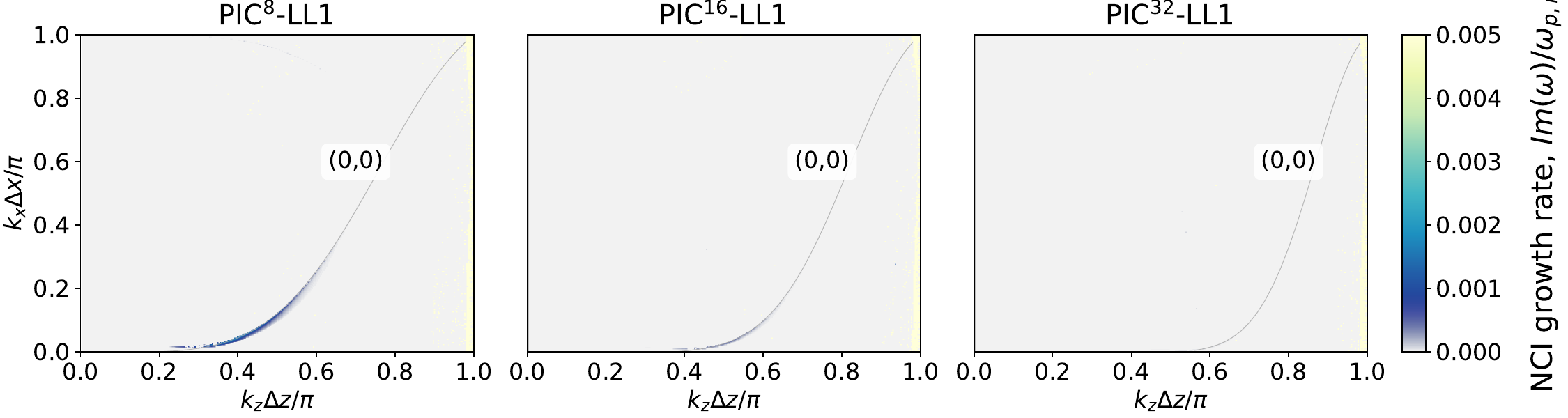}
\caption{\textbf{NCI growth rates}. Normalized NCI growth rates  $\text{Im}(\omega)/\omega_{p,r}$ in spectral space $(k_x,k_z)$, calculated from the analytical dispersion equation of the PIC$^p$-LL1 algorithm with different stencil order $p=8,16,32$. Solid grey lines correspond to $(m_t, m_z) = (0,0)$ mode, which blue-shifts as the stencil order increases. All numerical and physical parameters are the same as the ones used for the results reported in Figure~\ref{fig:unif_plasma_finite_order_small}. 
}
\label{fig:nci_gr_finite_order}
\end{figure}
\end{enumerate}

\subsection{
Laser-plasma acceleration}
This section demonstrates the extension of the stability properties observed in the uniform plasma cases to realistic 3D simulations of laser wakefield acceleration (LWFA)~\cite{Tajimaprl79}.
In these runs, a Gaussian laser pulse with amplitude $a_0=1.7$, duration $\tau = 73.3 \, \text{fs}$ and waist $w_0 =50 \, \mu\text{m}$ is injected at the entrance of a parabolic plasma channel with a background density $n_0=10^{18} \, \text{cm}^{-3}$ on axis.
The simulations are run in a Lorentz boosted frame of reference~\cite{Vayprl07} with $\gamma_0 = 60$ using the novel PIC$^{16}$-JR$_{ho}$m scheme (stencil order $p=16$ in all directions) with a hybrid alternating nodal-staggered grids (using field and current centering of order 16 in all directions) \cite{ZoniCPC2022}.
Similarly to the uniform plasma case, a bilinear filter was applied to the current and charge densities at each time step, with 4 passes in the $z$ direction and $1$ pass in the other directions.
The simulations were run on the Oak Ridge Leadership Computing Facility (OLCF) supercomputer Summit using 24 nodes (144 GPUs), with domain decomposition along both $x$ and $z$, using 24 guard cells in each direction.
\begin{figure}
\centering
\begin{subfigure}[b]{\textwidth}
\centering
\caption{$c\Delta t = \Delta x  = \Delta z/6$}
\includegraphics[width=\linewidth,trim={2cm 0.5cm 2cm 0.5cm},clip]{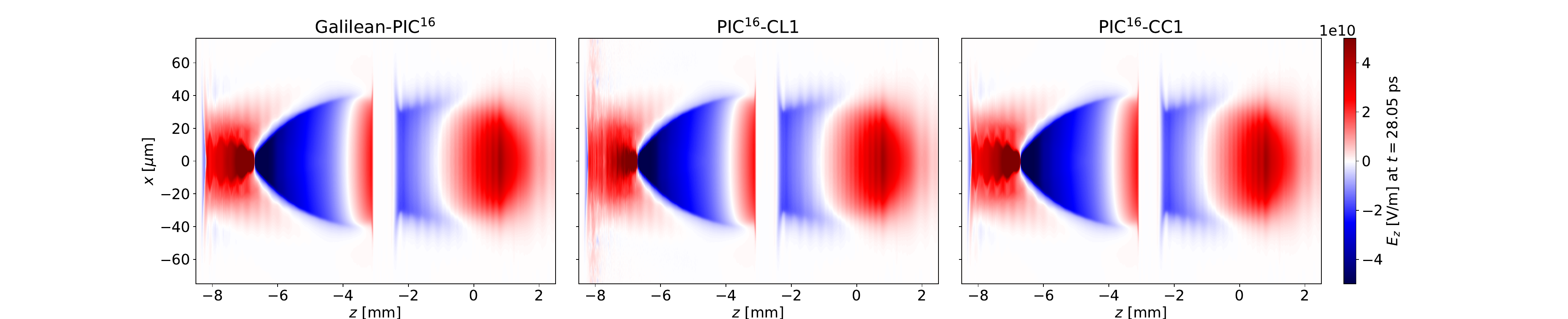}
\end{subfigure}
\begin{subfigure}[b]{\textwidth}
\centering
\caption{$c\Delta t = 3\Delta x  = \Delta z/2$}
\includegraphics[width=\linewidth,trim={2cm 0.5cm 2cm 0.5cm},clip]{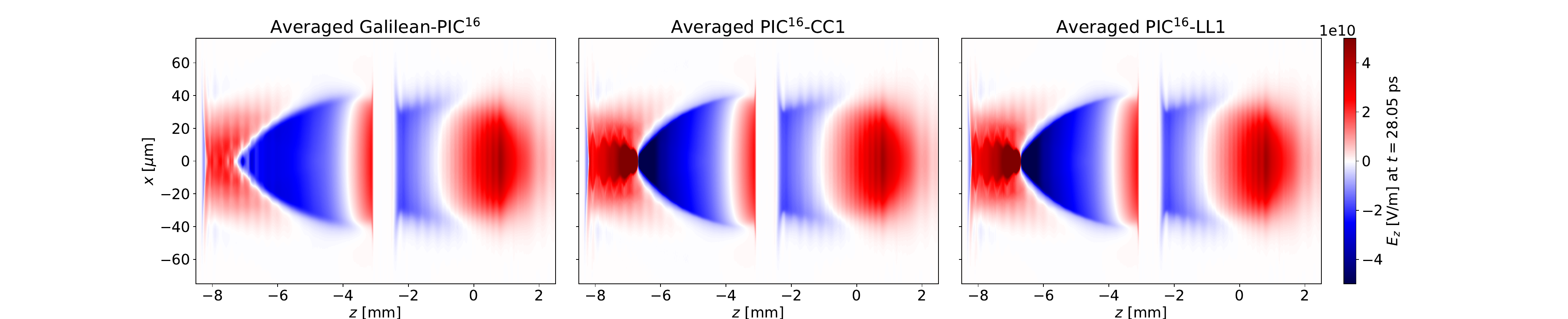}
\end{subfigure}
\begin{subfigure}[b]{\textwidth}
\centering
\caption{$c\Delta t = 6\Delta x  = \Delta z$}
\includegraphics[width=\linewidth,trim={2cm 0.5cm 2cm 0.5cm},clip]{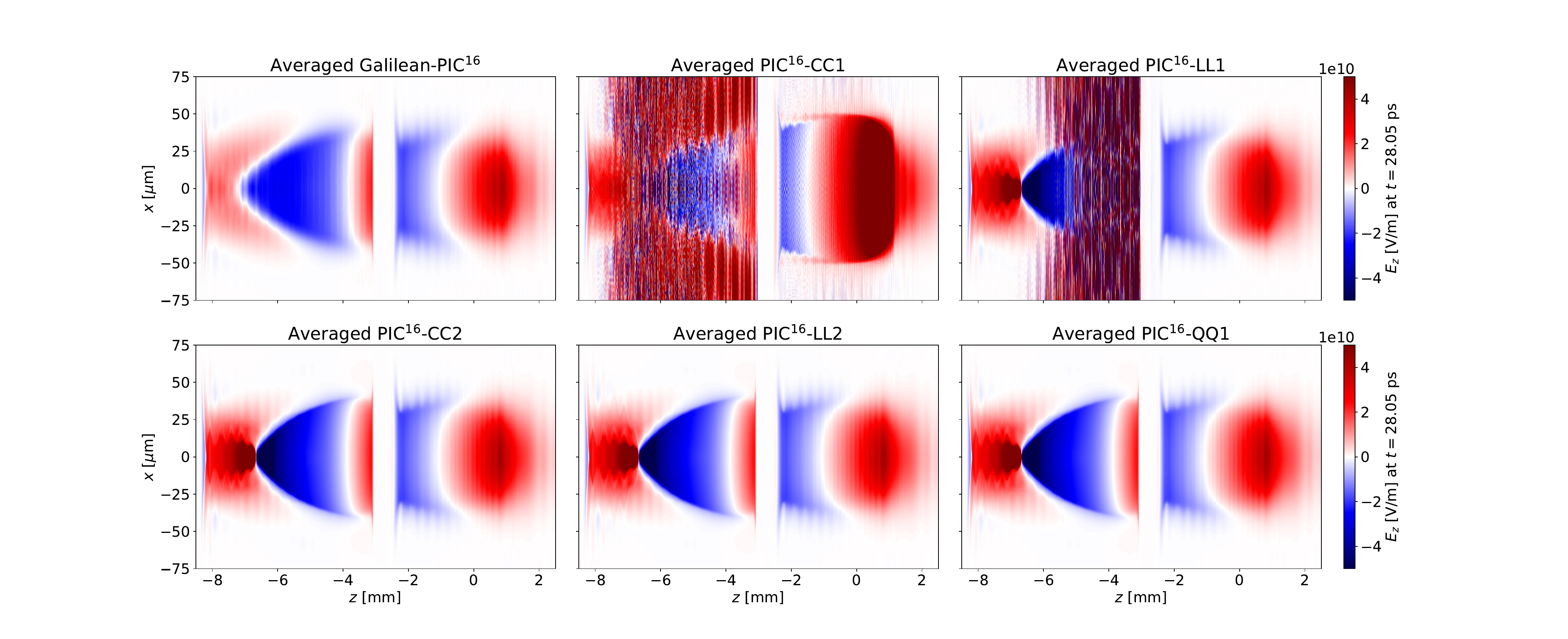}
\end{subfigure}
\caption{\textbf{WarpX simulation of LWFA with various time steps}. Snapshots of the longitudinal electric field $E_z$ $(x,z)$ slice at time $t=28.05 \, \text{ps}$ from the 3D simulation of two consecutive laser-driven plasma accelerator stages using the Galilean PIC$^{16}$ and PIC$^{16}$-JR$_{ho}$m (with JR$_{ho}$m = CC1, CL1, LL1, CC2, LL2 or QQ1) algorithms with time step (a) $c\Delta t = \Delta x = \Delta z/6$, (b) $c\Delta t = 3 \Delta x = \Delta z/2$ and (c) $c\Delta t = 6\Delta x  = \Delta z$. The laser (not shown) that drives the wake propagates from left to right.}
\label{fig:ez_3d_lwfa_JRm}
\end{figure}
%
%
\begin{figure}
\centering
\includegraphics[width=\linewidth]{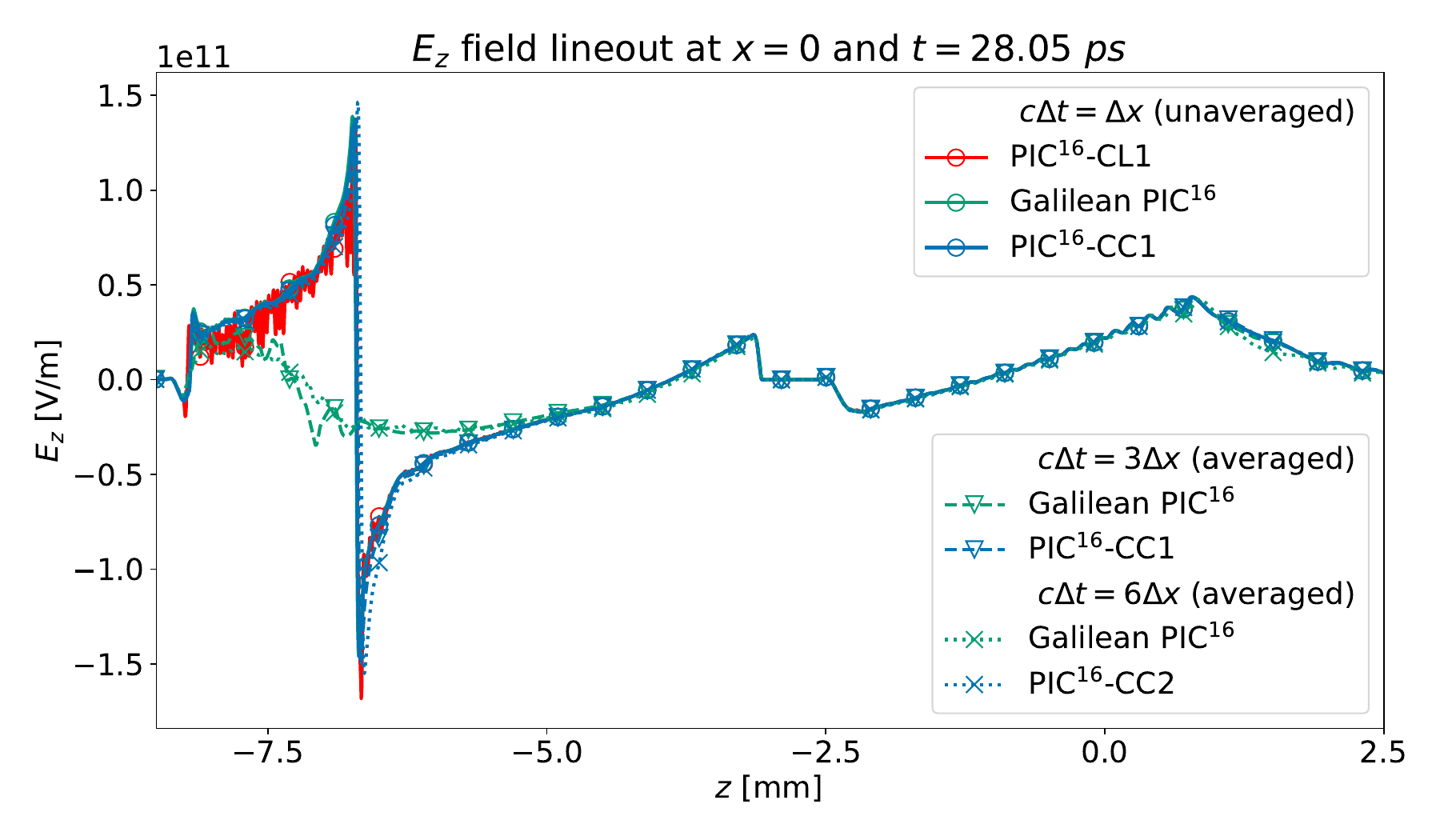}
\caption{\textbf{WarpX simulation of LWFA}. $E_z$-field lineouts at $x=0$ for selected cases of the results reported in Figure~\ref{fig:ez_3d_lwfa_JRm}.
}
\label{fig:3d_lwfa_simulation_lineout}
\end{figure}
The longitudinal resolution (in the boosted frame) was set to $\Delta z = (1+\beta_0)\gamma_0 \lambda_{lab} /24 = 4.08 \, \mu\text{m}$, where $\beta_0=\sqrt{1-1/\gamma_0^2}$ and $\lambda_{lab}=0.8 \, \mu\text{m}$ is the driving laser wavelength in the laboratory frame, while the transverse
resolution was $\Delta x = 0.68 \, \mu\text{m}$, so that $\Delta z = 6 \Delta x$. 
Simulations were also performed with the 
standard and averaged Galilean PIC$^{16}$ algorithm~\cite{LehePRE2016,KirchenPoP2016} 
for reference. 
%

Figure~\ref{fig:ez_3d_lwfa_JRm} displays snapshots of the longitudinal electric field $E_z$ 
from simulations running the Galilean PIC$^{16}$ and the PIC$^{16}$-JR$_{ho}$ algorithms at time $t  =28.05$~ps (which corresponds to $\omega_{p,r}t=84.3$) with different simulation time steps: (a) $c\Delta t=\Delta x=\Delta z/6$, (b) $c\Delta t=3\Delta x=\Delta z/2$ and (c) $c\Delta t=6 \Delta x=\Delta z$.
Figure \ref{fig:3d_lwfa_simulation_lineout} shows the corresponding lineouts at $x=0$ for a selection of runs.
Table~\ref{table_avg_runtimes} compares the performance of the various runs in each case.

When using the ``small'' time step $c\Delta t = \Delta x = \Delta z/6$, the PIC$^{16}$-CC1 algorithm is as effective as the standard Galilean PIC$^{16}$ algorithm for mitigating the NCI instability (which is emerging at the end of the second stage in the simulations using PIC$^{16}$-CL1),  with around 20\% speedup.
For larger time steps $c\Delta t = 3\Delta x = \Delta z/2$ and $c\Delta t = 6\Delta x = \Delta z$, 
although the averaged Galilean PIC$^{16}$ method is stable, it does not produce accurate physics results, leading to a very diminished amplitude of the electric field in the second stage.
Instead, the novel averaged PIC$^{16}$-JR$_{ho}$m method is stable and produces accurate results provided that the numbers of deposition and the number of timestep subintervals are high enough. 
For $c\Delta t = 3\Delta x = \Delta z/2$, both PIC$^{16}$-CC1 and PIC$^{16}$-LL1 are stable and accurate, with respective speedups of approximately $1.9$x and $1.5$x as compared to the Galilean reference case with small time steps. 
For $c\Delta t = 6\Delta x = \Delta z$, both PIC$^{16}$-CC1 and PIC$^{16}$-LL1 are unstable, while PIC$^{16}$-CC2, PIC$^{16}$-LL2 and PIC$^{16}$-QQ1 are stable and accurate, with respective speedups of approximately $2.1$x, $1.7$x and $1.8$x as compared to the Galilean reference case with small time steps.

%
%
%
%
\begin{table}[t]
\centering
\setlength{\tabcolsep}{5pt}
\begin{tabular}{c|ccccc}
\toprule
\makecell[c]{PSATD PIC solver} &$c\Delta t/\Delta x$& \makecell[c]{Averaged \\[-8pt]in time}& Stability  & \makecell[c]{Average time \\[-8pt] per step [s]} & \makecell[c]{Total time \\[-8pt] at $t=1.33 \, \text{ps}$ [s]} \\
\midrule
Galilean PIC &1& no& stable&0.465 & 279.27 \\
PIC-CL1 &1& no&unstable&0.44 &  261.67 \\
PIC-CC1 &1& no&stable&0.39 & 231.15 \\
\midrule
Galilean PIC &3&yes&\makecell[c]{stable, but \\[-8pt]inaccurate}&  0.54 & 109.6 \\
PIC-CC1 & 3&yes& stable & 0.7 & 144.83 \\
PIC-LL1 & 3&yes& stable & 0.9 & 182.9 \\
\midrule
Galilean PIC &6&yes&\makecell[c]{stable, but \\[-8pt]inaccurate}&  0.574 & 62.68 \\
PIC-CC1&6&yes &unstable&   0.83 &88.39 \\
PIC-CC2&6&yes &stable&   1.24 & 130.64 \\
PIC-LL1&6&yes & unstable  & 1.05 & 111.545 \\
PIC-LL2 &6&yes& stable &  1.61 & 165.4 \\
PIC-QQ1&6&yes & stable&  1.47 & 152.23 \\
\bottomrule
\end{tabular}
\caption{Performance comparison of runtimes for 3D LWFA simulations shown in Figure~\ref{fig:ez_3d_lwfa_JRm} using different spectral PIC solvers, run on the Summit supercomputer without I/O, using 24 nodes (144 GPUs), with domain decomposition in $x$ and $z$ and with 24 guard cells in each direction. Average time per step is from running up to time $t=1.33 \, \text{ps}$ which corresponds to the first 600 time steps with $c\Delta t = \Delta x$, or first 300 time steps with $c\Delta t = 3\Delta x$, or the first 100 time steps $c\Delta t = 6\Delta x$.}
\label{table_avg_runtimes}
\end{table}

These results show that the PIC$^p$-JR$_{ho}$m method is effective, efficient and versatile for controlling the numerical Cherenkov instability in plasma accelerator simulations, both in cases for which other methods (e.g., averaged Galilean PIC) apply as well, and in other cases that happen to be more challenging for the other methods.

\section{
\label{sec:conclusions}
Conclusion}
A novel formulation of pseudo-spectral analytical time-domain Particle-In-Cell algorithm is proposed and analyzed.
The formulation includes an additional term of ``hyperbolic divergence cleaning'' and a relaxation of the standard assumption of constant time dependency of the current density over one time step.
Extensions of the algorithm to finite-order stencils, alternating nodal-staggered grids and time-averaging over a time step were also presented.

Tests and analyses revealed that assuming the same time dependency for the evolution of the charge and current densities over one time step leads to excellent stability with regard to the numerical Cherenkov instability. Detailed analysis of the dispersion relation of the new algorithm (see Appendix \ref{sec:Appendix_dispeq}) provides a hint that explains the stability. 

The new algorithm is found to be effective, efficient and versatile for controlling the numerical Cherenkov instability in plasma accelerator simulations, both in cases for which other methods (e.g., Galilean PIC) apply and, more importantly, in other cases that happen to be more challenging for the other methods.
A possible extension of the algorithm for this particular application could be to incorporate the Galilean PIC algorithm in each subinterval, which should provide enhanced stability while preserving the versatility of the new scheme. 

While the application of the new algorithm to the modeling of plasma acceleration as proven successful, the application to other domains must be explored with care. For example, initial testings of the application of the method to the modeling of relativistic plasma shocks \cite{SpitkovskyAJ2008} has led to the observation of unphysical effects that have been tentatively attributed to unphysical coupling between the unphysical longitudinal electric field waves associated with divergence cleaning (from the term $F$ in Eq.~\ref{dF_dt}) and the plasmas. Further studies are needed to fully understand the underlying mechanisms and propose possible remedies. 

\appendix
\counterwithin{equation}{section}

\section{
\label{sec:Appendix_derivation_from_potentials}
Connection between the modified system of Maxwell's equations and a potential formulation}
\numberwithin{equation}{section}

It is instructive to derive the modified system of Maxwell's equations ~\eqref{Maxwell} in its potential form, starting with 

\begin{subequations}
\label{Maxwell_with_divb}
\begin{align}
\label{A:dE_dt}
\frac{\de\Eb}{\de{t}} & = c^2\Nabla\times\Bb-\frac{\Jb}{\eps_0}+c^2\Nabla{F} \,, \\[5pt]
\label{A:dB_dt}
\frac{\de\Bb}{\de{t}} & = -\Nabla\times\Eb \,, \\[5pt] \label{A:dF_dt}
\frac{\de F}{\de{t}} & = \Nabla\cdot\Eb-\frac{\rho}{\eps_0} \,, \\[5pt]
\label{A:divB}
\Nabla\cdot\Bb &  = 0\,.
\end{align}
\end{subequations}

Eq.~\eqref{A:divB} implies that $\Bb$ can be derived from a potential $\Bb=\Nabla\times\Ab$ which, when inserted in Eq.~\eqref{A:dB_dt}, gives $\Nabla\times(\Eb+\frac{\de\Ab}{\de t})=0$. 
This means that $\Eb+\frac{\de\Ab}{\de t}$ can be written as the gradient of a potential $\Phi$, giving
%
\begin{align}
\label{A:defE}
\Eb = -\Nabla\Phi - \frac{\de\Ab}{\de t} \,  .
\end{align}
%
Plugging \eqref{A:defE} in \eqref{A:dE_dt} and \eqref{A:dF_dt} leads to 
%
\begin{subequations}
\label{A:pot1}
\begin{align}
\label{A:potphi1}
\Nabla^2 \Phi +\frac{\de}{\de t} (\Nabla\cdot\Ab)
&  = -\frac{\rho}{\varepsilon_0}-\frac{\de F}{\de t}  \,, \\
\label{A:potA1}
\Nabla^2 \Ab - \frac{\de^2\Ab}{c^2\de t^2} 
 - \Nabla \left(\Nabla \cdot\Ab + \frac{1}{c^2}\frac{\de \Phi}{\de t}\right) 
 & = - \mu_0 \Jb + \Nabla F
 \,,
\end{align}
\end{subequations}
%
which, choosing $\Phi$ and $\Ab$ that verify the Lorentz gauge $\Nabla \Ab + \frac{1}{c^2}\frac{\de \Phi}{\de t}=0$, gives 
%
\begin{subequations}
\label{A:pot2}
\begin{align}
\label{A:potphi2}
\Nabla^2 \Phi -\frac{\de^2\Phi}{c^2 \de t^2} 
&  = -\frac{\rho}{\varepsilon_0}-\frac{\de F}{\de t}  \,, \\
\label{A:potA2}
 \Nabla^2 \Ab - \frac{\de^2\Ab}{c^2 \de t^2} 
 & = - \mu_0\Jb + \Nabla F \,, \\
\label{A:potF2}
\Nabla^2 F -\frac{\de^2 F}{c^2 \de t^2} 
&  = \mu_0 \left(\Nabla \cdot \Jb+\frac{\de \rho}{\de t}\right)  
 \,. 
\end{align}
\end{subequations}

A gauge transformation
%
\begin{subequations}
\label{A:gauge_Transf}
\begin{align}
\Ab' &= \Ab-\Nabla \Lambda, \\
\phi' &= \phi+\frac{1}{c^2}\frac{\de \Lambda}{\de t},
\end{align}
\end{subequations}
%
with
%
\begin{subequations}
\label{A:Fgauge}
\begin{align}
F &= \left(-\Nabla^2+\frac{1}{c^2}\frac{\de^2}{\de t^2}\right)\Lambda 
  = -\left(\Nabla \Ab' + \frac{\de \Phi'}{c^2 \de t}\right)
\end{align}
\end{subequations}
%
then leads to 
%
\begin{subequations}
\label{A:pot3}
\begin{align}
\label{A:potphi3}
\Nabla^2 \Phi' -\frac{\de^2\Phi'}{c^2 \de t^2} 
&  = -\frac{\rho}{\varepsilon_0}  \,, \\
\label{A:potA3}
 \Nabla^2 \Ab' - \frac{\de^2\Ab'}{c^2 \de t^2} 
 & = - \mu_0\Jb \,.
\end{align}
\end{subequations}

This is consistent with the derivation given in~\cite{Vay1996}, where Eqs.~\eqref{A:dE_dt}-~\eqref{A:dF_dt} were derived from Maxwell's equations in the Lorentz gauge form (i.e., the form of ~\eqref{A:potphi3}-~\eqref{A:potA3}) with the assumption that $\Jb=\Jb_0+\delta \Jb$ where $\Jb_0$ is the portion of $\Jb$ that verifies the continuity equation $\frac{\de \rho}{\de{t}}+\Nabla \Jb_0=0$, and defining $F=-\Nabla \delta \Ab$ such that $\Ab'=\Ab+\delta \Ab$ with $\Nabla \Ab + \frac{\de \Phi'}{c^2 \de t}=0$. 

In addition to showing that the term $F$ can arise from considerations other than a ``divergence cleaning'' term, this derivation also highlights how $F$ relates more directly to the continuity equation via Eq.~\eqref{A:potF2} and gauges via Eq.~\eqref{A:Fgauge}.

\section{
\label{sec:Appendix_updata_eqs}
Derivation of the PIC-JR\texorpdfstring{$_{ho}$}{ho}m equations}
\numberwithin{equation}{section}
We first rewrite Eqs.~\eqref{Maxwell_k} in an equivalent second-order differential form,

\begin{subequations}
\label{eq:Maxwell_k_2nd_or}
\begin{align}
\frac{\partial^2 \wh\Eb}{\partial t^2} +  c^2 k^2 \wh\Eb &= -\frac{1}{\varepsilon_0} \left( \frac{\partial \wh \Jb}{\partial t} + i c^2 \kb \wh\rho \right) \, \\
\frac{\partial^2 \wh\Bb}{\partial t^2} +  c^2 k^2 \wh\Bb &= \frac{1}{\varepsilon_0} i \kb \times \wh\Jb\, \\
\label{eq:B_Maxwell_k_2nd_or}
\frac{\partial^2 \wh F}{\partial t^2} + c^2 k^2 \wh F &= -\frac{1}{\varepsilon_0} \left( \frac{\partial \wh\rho}{\partial t} +  \kb \wh\Jb \right)\, 
%
\end{align}
\end{subequations}
%
and then sequentially integrate them analytically over each subinterval $[ n\Delta t + \ell \delta t, n\Delta t + (\ell+1)\delta t]$, $\ell \in [0,m-1]$ with $\delta t=\Delta t/m$, assuming that the current and charge densities are piecewise functions of time, given by Eqs.~(\ref{eq:j_func})-(\ref{eq:rho_func}).
Each of those equations can be expressed in the following generalized form with a right part as time polynomial up to order two:
\begin{align}
\begin{split}
\label{eq:Maxwell_k_2nd_or_generalized}
\Big( \frac{\partial^2}{\partial t^2} +  c^2 k^2\Big) \wh f &= \sum_{j=0}^{2}a_{0j} t^j \, \\
\end{split}
\end{align}
%
where $\{a_{0j}\}_{j=0}^{2}$ are known coefficients for any given $\wh f = \wh \Eb, \wh \Bb, \wh F$.
The general solution of such a second-order PDE equation with constant coefficients is:
\begin{align}
\begin{split}
\label{eq:2nd_or_pde_general_sol}
\wh f(t) &= C_{1} \cos(\omega(t-t_{n+\ell/m})) +  C_{2} \sin(\omega(t-t_{n+\ell/m})) \\
&+ \frac{1}{\omega^2} \Big( C_{3} (t-t_{n+(\ell+1/2){m}})^2 +  C_{4}(t-t_{n+(\ell+1/2){m}}) + C_{5}\Big),
\end{split}
\end{align}
%
where $\{C_k\}_{k=1}^{5}$ are integration coefficients to be determined. The coefficients $C_k$ with indexes $k=3,4,5$ for any given $\wh f = \wh \Eb, \wh \Bb, \wh F$ can be determined by solving a system of linear equations, obtained from substitution of Eq.~(\ref{eq:2nd_or_pde_general_sol}) into the corresponding Eq.~(\ref{eq:Maxwell_k_2nd_or}) and calculated at time steps $t_{n+\ell/m}$, $t_{n+(\ell+1/2)/m}$ and $t_{n+(\ell+1)/m}$. 
\begin{table}[t!]
\caption{Integration coefficients over $\ell$-th time subinterval $[n\Delta t + \ell \delta t,n\Delta t + (\ell+1) \delta t]$.}
\setlength{\tabcolsep}{10pt}
\begin{tabular}{ccccc}
\toprule
 $\wh f$& $\partial_{t} \wh f (t_n+\ell/m)$ & $C_3$ & $C_4$ & $C_5$ \\
  \hline
 $\wh \Eb$ & $ic^2 \kb \times \Bb^{n+\ell/m} -\frac{\Jb^{n+\ell/m}}{\varepsilon_0} $ & $-i\frac{2a_{\rho}^{\tau}c^2\kb}{\varepsilon_0 \delta t^2}$ & $-\frac{4 a_{\Jb}^{\tau} + i b_{\rho}^{\tau}c^2 \delta t \kb}{\varepsilon_0 \delta t^2}$ & $\frac{4 i a_{\rho}^{\tau} c^2 \kb - b_{\Jb}^{\tau}\omega^2 \delta t- i c^2 c_{\rho} \omega^2 \delta t^2 \kb} {\varepsilon_0 \delta t^2}$\\ %
 & $+ ic^2 \wh F^{n+\ell/m}\kb$ &  & & \\ %
 $\wh \Bb$& $-i\kb \times \Eb^{n+\ell/m}$ & $i\frac{2a_{\Jb}^{\tau}c^2\kb \times}{\varepsilon_0 \delta t^2}$ & $i\frac{b_{\Jb}^{\tau} \kb \times}{\varepsilon_0 \delta t}$ & $\frac{-4 i a_{\Jb}^{\tau} \kb \times + i c_{\Jb} \omega^2 \delta t^2 \kb \times } {\varepsilon_0 \delta t^2 \omega^2}$ \\
$\wh F$ & $i\kb\cdot\wh\Eb^{n+\ell/m}-\frac{\wh\rho^{n+\ell/m}}{\eps_0}$ & $-i\frac{2a_{\Jb}^{\tau}\kb }{\varepsilon_0 \delta t^2}$ & $-i\frac{4a_{\rho}^{\tau}\kb + i b_{\Jb} \delta t \kb}{\varepsilon_0 \delta t^2}$& $\frac{4 i a_{\Jb}^{\tau} \kb  - b_{\rho} \delta t \omega^2 - i c_{\Jb} \omega^2 \delta t^2 \kb } {\varepsilon_0 \delta t^2 \omega^2}$ \\
\toprule
\end{tabular}
\end{table}
While the remaining coefficients $C_{1}$ and $C_2$ can be determined from the initial conditions $\wh f(t)|_{t_{n+\ell/m}}$ and $\partial_{t} \wh f(t)|_{t_{n+\ell/m}}$, respectively, \\
\begin{align}
\begin{split}
\label{eq:C1_C2_coefs}
C_1 &= f(t_n+\ell/m) - \Big( C_3 (\delta t/2)^2 + C_4 (\delta t /2) + C_5 \Big)/\omega^2 ,\ \\
C_2 &=  \partial_t f(t_n+\ell/m) - \Big( 2 C_3 (\delta t/2)+C_4 \Big) / \omega^2.
\end{split}
\end{align}
The expression of the field components $\wh f (t_{n+(\ell+1)/{m}})$ at the next time subinterval are then given by:
\begin{align}
\begin{split}
\label{eq:2nd_or_pde_general_sol_at_next_tstep}
\wh f (t_{n+(\ell+1)/m}) = C_{1} \cos(\omega \delta t ) +  C_{2} \sin(\omega \delta t) + \frac{1}{\omega^2} \Big( C_{3} ( \delta t /2) ^2 +  C_{4} (\delta t / 2)+ C_{5}\Big). \
\end{split}
\end{align}
\section{
\label{sec:Appendix_averaged_eqs}
Derivation of the averaged PIC-JR\texorpdfstring{$_{ho}$}{ho}m equations}
\numberwithin{equation}{section}
The notation $\langle \wh f(t)\rangle^{n+1}$ is introduced to refer to the average of any given function $\wh f(t)$ over the time interval $[n\Delta t, (n+1)\Delta t]$ as,
\begin{equation}
\langle\wh{f}\rangle^{n+1} = \frac{1}{2 \Delta t} \int_{t_n}^{t_n+2\Delta t} {\wh f(t') \dd t'} , \ \text{where}\  \wh f = \wh \Eb, \wh \Bb.
\label{eq:EB_avg}
\end{equation}
%
For any given number of subintervals $m$, the integral in Eq.(~\ref{eq:EB_avg}) can be split into a sum over $2m$ integrals over $[{t_n}+\ell \delta t, {t_n} + (\ell+1)\delta t], \ \ell=0,..,2m-1$ as,
\begin{equation}
\label{eq:f_avg}
\langle\wh{f}\rangle^{n+1} = \frac{1}{2 \Delta t} \sum_{\ell=0}^{2m-1}{\int_{t_n +\ell \delta t}^{t_n+(\ell+1)\Delta t} {\wh f(t') \dd t'}}, \ \text{where}\  \wh f = \wh \Eb, \wh \Bb.
\end{equation}
%
The averaged $\langle \wh \Eb \rangle$ and $\langle \wh \Bb \rangle$ fields are obtained through sequential integration of Eq.~(\ref{eq:2nd_or_pde_general_sol}) over each subinterval  $[t_n+\ell\delta t, t_n+(\ell+1)\delta t], \ \ell=0,..,2m-1$ and then susbstituted into Eq.~(\ref{eq:f_avg}),
\begin{subequations}
\begin{align}
\int_{t_n+\ell\delta t }^{t_{n}+(\ell+1)\delta t} {\wh \Eb(t') \dd t'} &=  \frac{S}{ ck } \wh{\Eb}^{n+\ell/m} +  \frac{i c^2 Y_4}{c^2k^2} \kb \times \wh \Bb^{n+\ell/m}  + \frac{i Y_4}{2ck\delta t} \wh{F}^{n+\ell/m}  \nonumber  \\
& + \frac{1}{\varepsilon_0c^2k^2}(  Y_1 \ab_j - Y_5  \bb_J  - Y_4  \cb_J )  - i c^2 \kb \times (  Y_6 a_{\rho} + Y_7 b_{\rho}  + Y_8  c_{\rho} ) \,, \\
\int_{t_n+\ell\delta t }^{t_{n}+(\ell+1)\delta t} {\wh \Bb(t') \dd t'}  &= \frac{S}{ck}\wh{\Bb}^{n+\ell/m} - \frac{i Y_4}{c^2k^2} \kb \times \wh \Eb^{n+\ell/m}  + i \kb \times (  Y_6 \ab_J + Y_7 \bb_J  + Y_8  \cb_J ) \,,
\end{align}
\end{subequations}
with
\begin{subequations}
\begin{align}
Y_{6} &= \frac{1}{6 \varepsilon_0 c^5 k^5 \delta t^2} \left((ck\delta t)^2 - 3 \delta (ck\delta t)^2 S -12 ck \delta t (1+C)+24 S \right) \,, \\
Y_{7} &= \frac{1}{2 \varepsilon_0 c^4 k^4 \delta t} \left(ck\delta t S + 2 (C-1) \right) \,, \\
Y_{8} &= \frac{\delta t}{\varepsilon_0 c^2 k^2} \left(1 - \frac{S}{ck\delta t}\right) \,.
\end{align}
\end{subequations}
%
\section{
\label{sec:Appendix_dispeq}
Dispersion relation for the PIC-JR\texorpdfstring{$_{ho}$}{ho}m algorithm}
\numberwithin{equation}{section}
\numberwithin{equation}{section}
The 2D dispersion relation for Eqs.~\eqref{Maxwell_discrete} is derived to analyze the algorithm's stability with respect to the numerical Cherenkov instability (NCI), for a uniform plasma flowing through a periodic grid along the $z$-axis with a velocity $\vb_0 = (0,0,v_0)$, where $v_0 = c ( 1 - 1/\gamma_0^2 )^{1/2}$.
%
%
Following the analysis from \cite{ShapovalPRE2021, LehePRE2016}, we consider the discretized perturbed Vlasov equation, expressed in Fourier space:
\begin{equation}
\label{eq:supp-vlasov}
\begin{split}
\delta \hat{f}^{n+1/2} & ({\kb_m},\pb)e^{i{\kb_m}\cdot\vb\Delta t/2} \\
& -\hat{f}^{n-1/2}(\kb_m,\pb)e^{-i\kb_m\cdot\vb\Delta t/2} \\
& +q\Delta t \hat{S}(\kb_m)\Big[\spectral{\Eb}^n({\kb}) + \vb\times \spectral{\Bb}^n(\kb) \Big] \cdot \frac{\partial f_0}{\partial \pb} = 0 \,,
\end{split}
\end{equation}
where $f_0 =n_0 \delta (\pb - m\gamma_0\vb_0)$ is the distribution function of the uniform plasma in a state of equilibrium, and $\delta f$ is a perturbation to $f_0$.
The discretized formulas for the deposited current and charge in Fourier space at any time $\ell\Delta t$, $\ell \in [n,n+1]$ centered around $\delta \hat{f}^{n+1/2}$, are given by
\begin{align}
\label{eq:supp-J-depos}
\wh{\Jb}^{\ell}(\kb) &= \sum_{m}{S(\kb_m)}\int{d\pb q\vb\delta \hat{f}^{n+1/2}(\kb_m,\pb)} e^{-i \kb_m\cdot\vb (\ell-(n+1/2))\Delta t}, \\[5pt]
\label{eq:supp-rho-depos}
\wh{\rho}^{\ell}(\kb) &= \sum_{m}{S(\kb_m)}\int{d\pb q \delta \hat{f}^{n+1/2}(\kb_m,\pb)} e^{-i \kb_m\cdot\vb (\ell-(n+1/2))\Delta t}.
\end{align}
Then, assuming the same $e^{-i\omega t}$ time evolution for $\wh \Eb, \wh \Bb, \wh F, \wh \Jb, \wh \rho$ and $\wh \delta  f$ with the following anzatz,
\begin{subequations}
\label{eq:supp-modes}
\begin{align}
\wh{\Eb}^{n}({\kb}) & = \wh{\Eb}({\kb}) e^{-i\omega n\Delta t} \,, \\
\delta \hat{f}^{n+1/2}({\kb_m},\pb) & =\delta \hat{f}({\kb_m},{\pb}) e^{-i\omega (n+1/2)\Delta t} \,, \\
{ \wh \Jb}^{n}({\kb}) & = \wh{\Jb}({\kb}) e^{-i\omega  n\Delta t} \,, \\
{\wh{\rho}}^{n}({\kb}) & = \wh{\rho}({\kb}) e^{-i\omega n\Delta t} \,.
\end{align}
\end{subequations}
equation~\eqref{eq:supp-vlasov} yields
\begin{equation}
\label{eq:supp-dicrete_df}
\begin{split}
\delta \hat{f}(\kb_m, \pb) &= -i \frac{q \Delta t}{2} \wh S(\kb_m)\frac{\wh \Eb(\kb)+\vb \times \wh \Bb(\kb)}{\sin((\omega-\kb_m\cdot\vb)\Delta t/2)}.
\end{split}
\end{equation}
Substituting the Vlasov equation~\eqref{eq:supp-vlasov} into \eqref{eq:supp-J-depos}-\eqref{eq:supp-rho-depos} gives the following expressions for the deposited current $\wh{\Jb}(\kb)$ and the charge $\wh \rho(\kb)$:
\begin{align}
\label{eq:supp-j}
\wh{\Jb} & = i\frac{ck \varepsilon_0}{\wh{T}} \left(\xi_0 +({\xib} \cdot  \wh{\Qb})\frac{ {\vb}}{c}\right) \,, \\
\label{eq:supp-rho}
{\wh \rho} & = \frac{i k\varepsilon_0 } {\wh{T}} ({\xib} \cdot  {\wh \Qb} ) \,,\\
\wh{\Qb}({\kb}) &=\wh{\Eb} ({\kb})+ {\vb} \times  \wh{\Bb}({\kb})  - ({\vb} \cdot   \wh{\Eb} ({\kb}) ) {\vb} / c^2 \,, \\
\label{eq:supp-j-ksi-coefs}
\xi_{0}&= \frac{\wh{T}\omega_p^2}{\gamma_0 ck} \sum_{m=-\infty}^{+\infty}S^2({\bf{k_m}}) \cdot \frac{1}{\frac{2}{\Delta t}s_{\omega}'}, \\
\xib&= \frac{\wh{T}\omega_p^2}{\gamma_0 k} \sum_{m=-\infty}^{+\infty}S^2({\bf{k_m}}) \cdot \frac{{\bf{k_m}} c_{\omega}'}{\Big(\frac{2}{\Delta t}s_{\omega}'\Big)^2} \,,
\end{align}
where $\wh{T} = \prod_{i} \big[ 1-\sin(k_i \Delta i/2) \big]$ is one pass of a  binomial smoothing operator, and $\omega_p = (n_0 q^2 m_e^{-1} \varepsilon_0^{-1})^{1/2}$ is the plasma frequency, and $\wh{S}({\kb_m})$ is the particle shape factor.
Still following \cite{ShapovalPRE2021,LehePRE2016}, Eqs. \eqref{Maxwell_discrete} are then rewritten in the time-symmetrical form
%
%
%
\begin{subequations}
\begin{align}
\label{eq:symm-maxwellE}
\left(\wh{\Eb}^{n+(\ell+1)/m} - \wh{\Eb}^{n+\ell/m}\right) &=  i\frac{S}{(1+C)ck} c^2 \kb \times \left(\wh\Bb^{n+(\ell+1)/m)} +\wh\Bb^{n+\ell/m}\right) \nonumber \\
&+ i \frac{S}{(1+C)ck} c^2 \kb \left( \wh{F}^{n+(\ell+1)/m} +  \wh{F}^{n+\ell/m}\right) \nonumber\\
&+ \frac{1}{\varepsilon_0 \omega} \Big( Y_{9} \ab_j - 2S (1+C)^{-1}\cb_j \Big) -\frac{i c^2}{\varepsilon_0 c^2k^2} Y_{10} \kb  b_{\rho}, \\
\label{eq:symm-maxwellB}
\left(\wh{\Bb}^{n+(\ell+1)/m} - \wh{\Bb}^{n+\ell/m}\right) &= - \frac{S}{(1+C)ck} i \kb \times \left(\wh\Eb^{n+(\ell+1)/m}+\wh\Eb^{n+\ell/m}\right) + \frac{i\kb  \times \bb_j }{\varepsilon_0 c^2 k^2}  Y_{10} \,, \\[5pt]
\label{eq:symm-maxwellF}
\left(\wh{F}^{n+(\ell+1)/m} - \wh{F}^{n+\ell/m}\right)  &= \frac{S}{(1+C)ck}i\kb \left(\wh{\Eb}^{n+(\ell+1)/m} + \wh{\Eb}^{n+\ell/m}\right) \nonumber \\
&- \frac{1}{\varepsilon_0 c^2 k^2} i \kb \bb_j Y_{10} + \frac{1}{\varepsilon_0 ck} \Big( Y_9 a_{\rho} -2S  (1+C)^{-1}c_{\rho}\Big) \,. 
\end{align}
\label{time_symm_maxwell}
\end{subequations}
%
%
%
%
Substitution of equations~\eqref{eq:supp-modes} in Eqs.\eqref{eq:symm-maxwellE}-\eqref{eq:symm-maxwellF} gives
\begin{subequations}
\begin{align}
\label{eq:supp_e_discret}
s_{\omega}\wh \Eb & = - t_{ck}c_{\omega} {\kb} \times c\wh\Bb - c_{\omega} t_{ck} \hat{\kb} c\wh F + i \Big (Y_9  \tilde{a}_{\omega}^{\tau} / 2 -  t_{ck} \tilde{c}_{\omega}^{\tau} \Big) \wh \Jb + \Big( Y_{10}\tilde{b}_{\omega}^{\tau} / 2 \Big)  \kb \wh \rho, \\
\label{eq:supp_b_discret}
s_{\omega}\wh \Bb & =  t_{ck} c_{\omega} \kb \times \wh \Eb - \Big(Y_{10} \tilde{b}_{\omega}^{\tau} / 2\Big) \kb \times \wh \Jb, \\
\label{eq:supp_f_discret}
s_{\omega} c\wh F & = - t_{ck} c_{\omega} {\kb} \cdot \wh \Eb +  \kb \cdot \wh \Jb \Big (Y_{10}  \tilde{b}_{\omega}^{\tau} / 2 \Big) + i \Big(  Y_9 \tilde{a}_{\omega}^{\tau} / 2 - t_{ck}  \tilde{c}_{\omega} \Big) \wh \rho \,.
\end{align}
\end{subequations}

Projecting Eqs.~\eqref{eq:supp-j} and \eqref{eq:supp_e_discret} along $x$ and $z$ and Eqs. \eqref{eq:supp_b_discret} along $y$ gives the following 2D dispersion equation in matrix form: 
\begin{subequations}
\begin{equation}
\label{eq:matrix_eq}
\Mb\Ub^{T} = 0 \,, \\
\end{equation}
\begin{equation}
\label{eq:matrix}
\fontsize{10.5}{11}\selectfont
\Mb =
\begin{bmatrix}
  & -s_{\omega}        &  0 & c_{\omega}\wh{k}_z t_{ck} & -c_{\omega}  \wh{k}_x t_{ck}         &i T {\color{black}{\chi_{\tau_{J}}}}& 0         &-iT \wh{k}_x {\color{black}{\psi_{\tau_{\rho}}}}   &  \\
  \\
  &0 &  -s_{\omega}  & -c_{\omega} \wh{k}_x t_{ck}        & -c_{\omega}  \wh{k}_z t_{ck} & 0         & i T {\color{black}{\chi_{\tau_{J}}}}     & -iT \wh{k}_z  {\color{black}{{\psi_{\tau}}_{\rho}}}&    \\
 \\
& c_{\omega} \wh{k}_z t_{ck}  & -c_{\omega} \wh{k}_x t_{ck} & -s_{\omega}  & 0& iT \wh{k}_z  {\color{black}{{\psi_{\tau_{J}}}}}& -iT \wh{k}_x  {\color{black}{\psi_{\tau_{J}}}}& 0& \\
\\
 & -c_{\omega} \wh{k}_x t_{ck}  & -c_{\omega} \wh{k}_z t_{ck} & 0& -s_{\omega}  &-iT \wh{k}_{x} {\color{black}{\psi_{\tau_{J}}}} &  -iT \wh{k_{z}} {\color{black}{\psi_{\tau_{J}}}} &i T {\color{black}{{\chi_{\tau}}_{\rho}} }& \\
 \\
 & \frac{i}{T}  \xi_0& 0& -\frac{i}{T} \xi_0\beta_0 & 0 & -1 & 0&0& \\
 \\
 &\frac{i}{T} \xi_x \beta_0 & \frac{i}{T} (1-\beta_0^2)(\xi_0 + \xi_z  \beta_0)& -\frac{i}{T} \xi_x \beta_0^2& 0&0&-1&0 &\\
 \\
 &\frac{i}{T}\xi_x & \frac{i}{T} \xi_z(1-\beta_0^2)& -\frac{i}{T} \xi_x \beta_0& 0&0&0 &-1&\\
\end{bmatrix}
\end{equation}
\end{subequations}
where $\Ub = \left( \wh{E}_x , \wh{E}_z, c\wh{B}_y, c\wh{F}, \wh{J}_x/(c k \, \varepsilon_0), \wh{J}_z/(c k \, \varepsilon_0), \wh{\rho}/(k \, \varepsilon_0)\right)$ and $\wh{\kb} = \kb/k$ is the normalized wave vector.
The matrix coefficients in $\Mb$ that depend on the time dependency of the current and charge densities $\wh \Jb$ and $\wh \rho$ are summarized in Table~\ref{table:matrix_coefs}. For example, the upper index $\tau_{J/\rho}$ in the coefficients $\psi_{\tau_{J/\rho}}$ and $\chi_{\tau_{J/\rho}}$ indicates the time dependency of $\wh \Jb$ and $\wh \rho$ and can be \textit{constant} (C), \textit{linear} (L) or \textit{quadratic} (Q).
\begin{table}[ht]
\caption{Matrix coefficients of the dispersion equation \eqref{eq:matrix_eq}, based on the time dependency of the current and charge densities $\wh \Jb$ and $\wh \rho$ over one time subinterval, $\delta t = \Delta t/m$.}
\setlength{\tabcolsep}{10pt}
\begin{tabular}{cccc}
\toprule
\multirow{2}{*}{Coefficients}& \multicolumn{3}{c}{Time dependency of $\wh \Jb$ or $\wh \rho$} \\
\cline{2-4}
& constant  ($\tau=0$) & linear  ($\tau=1$) & quadratic  ($\tau=2$) \\ 
\midrule
${\tilde{a}_{\omega}^{\tau}}$ & $0$ & $0$ & $(c_{\omega}-1)$ \\ 
${\tilde{b}_{\omega}^{\tau}}$ & $0$ & $s_{\omega}$ & $s_{\omega}$ \\
${\tilde{c}_{\omega}^{ \tau}}$ & $1$ & $c_{\omega}$ & $1$ \\
${\chi_{\tau}}$ & $-t_{ck}$ & $-c_{\omega} t_{ck}$ & $Y_9 (c_{\omega}-1)-t_{ck}$ \\
$\psi_{\tau}$ & $0$ & $-i s_{\omega} Y_{10}$ & $-i s_{\omega} Y_{10}$ \\
\bottomrule
\end{tabular}
\label{table:matrix_coefs}
\end{table}
%
%

The other coefficients are given by:
\begin{subequations}
\label{eq:matrix_coefs}
\begin{align}
c_{\omega} &= \cos(\omega \delta t/2), \\
s_{\omega} &= \sin(\omega \delta t/2), \\
t_{\omega} &= s_{\omega} /c_{\omega}, \\
c_{\omega}' &= \cos \left((\omega - \kb_m \cdot \vb) \Delta t/2 \right), \\
s_{\omega}' &= \sin \left((\omega-\kb_m \cdot \vb) \Delta t/2 \right), \\
\kb_m &= \kb + 2\pi m/\Delta\rb,\ m \in Z, \\
t_{ck} &= \tan(ck\delta t/2), \\
Y_9 &=\frac{ t_{ck} (8-c^2k^2\delta t^2)-4ck\delta t}{(1+C)(ck\delta t)^2}, \\
Y_{10} &= \left(1-\frac{2t_{ck}}{ck\delta t}\right), \\
\chi_{\tau} &=Y_9 \tilde{a}_{\omega}^{\tau} - t_{ck}   \tilde{c}_{\omega}^{\tau}, \\
\psi_{\tau} &= Y_{10} \tilde{b}_{\omega}^{\tau}.
\end{align}
\end{subequations}
The dispersion relation is given by computing the determinant of  $(\Mb)$ using the Sarrus rule.
Interestingly, when the charge and current densities have the same temporal dependency, e.g., with CC, LL or QQ, the determinant simplifies to the straightforward expression
\begin{equation}
\det(\Mb) = \alpha_1 \alpha_2, 
\end{equation}
where
\begin{subequations}
\label{eq:alfa_coefs}
\begin{align}
\alpha_1 &= \wh{T}^3 \Big [\xi_0 \Big (\beta_0 \wh{k}_z (\chi_{\tau} c_{\omega} t_{ck} + \psi_{\tau} s_\omega) - (\chi_{\tau} s_\omega + \psi_{\tau} c_\omega t_{ck})\Big ) + (c_\omega^2 t_{ck}^2-s_\omega^2)\Big ], \\
\alpha_2 &= (c_\omega^2 t_{ck}^2 - s_\omega^2)  + (1-\beta_0^2) \Big [(\xi_x \wh{k}_x + \xi_z \wh{k}_z) (\chi_{\tau} c_\omega t_{ck} + \psi_{\tau} s_\omega) \nonumber \\
& + \psi_{\tau} c_\omega t_{ck}(\xi_0 + \xi_z \beta_0) + \chi_{\tau} (\xi_z c_\omega \beta_0 + \xi_0 s_\omega)\Big].
\end{align}
\end{subequations}
%

Here, such simplification is possible due to the presence of similar terms of opposite sign that cancel each other when the charge and current densities have the same time dependency.
For example, terms like $\big(\psi_{\tau_J}\big)^2 k_x k_z c_\omega^2 -\psi_{\tau_J}\psi_{\tau_\rho}  k_x k_z c_\omega^2 = 0$, since $\psi_{\tau_J} = \psi_{\tau_{\rho}} = \psi_{\tau}$ ($\tau_{\Jb}= \tau_{\rho}=\tau$).
Moreover, at the asymptotic limit, assuming that (i) $\delta \omega = \omega-\kb_m \vb_0$ is small and (ii) considering an ultra-relativistic regime, e.g., $\beta_0=v_0/c=1$, 
the determinant equation reduces to:
\begin{align}
\xi_0 \Big ( \wh{k}_z (\chi_{\tau} c_{k_mv_0} t_{ck} + \psi_{\tau} s_{k_mv_0}) - (\chi_{\tau} s_{k_mv_0} + \psi_{\tau} c_{k_mv_0} t_{ck})\Big ) + (c_{k_mv_0}^2 t_{ck}^2-s_{k_mv_0}^2) = 0
\end{align}
where $c_{k_m v_0} = \cos(\kb_m \vb \delta t/2)$, $s_{k_m v_0} = \sin(\kb_m \vb \delta t/2)$, and $\xi_0^{\tau}$ is proportional to $1/\delta \omega$ and reads
\begin{align}
\xi_{0}^{\tau}&= \frac{\wh{T}\omega_p^2 S^2(\kb_m)}{\gamma_0 ck} \frac{1}{\delta \omega} + \frac{\wh{T}\omega_p^2}{\gamma_0 ck} \sum_{j=-\infty, \ m \neq j}^{+\infty}S^2({\bf{k_j}}) \cdot \frac{1}{\frac{2}{\Delta t}s_{k_j v_0}'} = \frac{\alpha_m}{\delta \omega} + \beta_m \,.
\end{align}
Finally, we obtain a first order equation for $\delta \omega$ with real coefficients,
\begin{align}
\delta \omega &= - \frac{\alpha_m \Big ( \wh{k}_z (\chi_{\tau} c_{k_mv_0} t_{ck} + \psi_{\tau} s_{k_mv_0}) - (\chi_{\tau} s_{k_mv_0} + \psi_{\tau} c_{k_mv_0} t_{ck})\Big ) }{\beta_m \Big ( \wh{k}_z (\chi_{\tau} c_{k_mv_0} t_{ck} + \psi_{\tau} s_{k_mv_0}) - (\chi_{\tau} s_{k_mv_0} + \psi_{\tau} c_{k_mv_0} t_{ck})\Big ) + (c_{k_mv_0}^2 t_{ck}^2-s_{k_mv_0}^2)} \,.
\end{align}
It follows that, under the assumptions (i)-(ii), the determinant has only real coefficients, $\delta\omega$ is real, and the algorithm is stable.
\begin{acknowledgments}
This research used the open-source particle-in-cell code WarpX (\url{https://github.com/ECP-WarpX/WarpX}).
We acknowledge all WarpX contributors.
This research was supported by the Exascale Computing Project (17-SC-20-SC), a collaborative effort of the U.S. Department of Energy Office of Science and the National Nuclear Security Administration.
This research was performed in part under the auspices of the U.S. Department of Energy by Lawrence Berkeley National Laboratory under Contract DE-AC02-05CH11231.
This research used resources of the Oak Ridge Leadership Computing Facility, which is a DOE Office of Science User Facility supported under Contract DE-AC05-00OR22725.
The data that support the findings of this study are available from the corresponding author upon reasonable request.
\end{acknowledgments}

Olga Shapoval derived and implemented the algorithms in their final forms, performed the numerical analyses and numerical tests. 
Edoardo Zoni contributed to the derivation, implementation and testing of the algorithms.
Rémi Lehe wrote the initial implementation of the algorithm in WarpX and discussed the results.
Maxence Thévenet performed numerical tests of an early prototype implemented in the code Warp.
Jean-Luc Vay proposed the concept and implemented an early prototype in the code Warp.

\bibliography{bibliography,library}
\end{document}